\documentclass[econometrics,article,accept,pdftex,moreauthors]{Definitions/mdpi}

\bibpunct{(}{)}{;}{a}{}{,~}
\firstpage{1} 
\pubvolume{1}
\issuenum{1}
\articlenumber{0}
\pubyear{2023}
\copyrightyear{2023}

\datereceived{27 November 2022} 
\daterevised{3 March 2023} 
\dateaccepted{3 March 2023} 
\datepublished{} 
\hreflink{https://doi.org/} 

\usepackage{amssymb,amsmath,mathrsfs,stmaryrd,amsthm,mathtools,amsfonts,graphicx,url,booktabs,nicefrac,comment,microtype,tikz-cd,lingmacros,nccmath,tree-dvips, bbm, bm, etoolbox,algorithm,caption,color,enumitem, physics, lineno, setspace} 

 \usepackage[labelformat=simple]{subcaption}

\DeclareCaptionLabelFormat{subcaptionlabel}{\normalfont(\textbf{#2}\normalfont)}
\DeclareMathOperator*{\argmax}{argmax}
\DeclareMathOperator{\E}{\mathbb{E}}
\DeclareMathOperator{\Cov}{\text{Cov}}

\Title{Semi-metric portfolio optimization: a new algorithm reducing simultaneous asset shocks} %

\TitleCitation{Semi-metric portfolio optimization: a new algorithm reducing simultaneous asset shocks}

\Author{Nick James $^{1,}$*$^{,\dagger}$\orcidA{}, Max Menzies $^{2,\dagger}$\orcidB{} and Jennifer Chan $^{3}$\orcidC{}}

\AuthorNames{Nick James, Max Menzies and Jennifer Chan}

\AuthorCitation{ James, Nick, Max Menzies, and Jennifer Chan}

\address{%
$^{1}$ \quad School of Mathematics and Statistics, University of Melbourne, Parkville, VIC 3010, Australia \\
$^{2}$ \quad Beijing Institute of Mathematical Sciences and Applications, 
Tsinghua University, Beijing 101408, China; max.menzies@alumni.harvard.edu \\
$^{3}$ \quad School of Mathematics and Statistics, University of Sydney, Camperdown, NSW 2006, Australia; jennifer.chan@sydney.edu.au
}

\corres{Correspondence: nick.james@unimelb.edu.au}

\firstnote{These authors contributed equally to this work.}

\abstract{This paper proposes a new method for financial portfolio optimization based on reducing simultaneous asset shocks across a collection of assets. This may be understood as an alternative approach to risk reduction in a portfolio based on a new mathematical quantity. First, we apply recently introduced semi-metrics between finite sets to determine the distance between time series' structural breaks. Then, we build on the classical portfolio optimization theory of Markowitz and use this distance between asset structural breaks for our penalty function, rather than portfolio variance. Our experiments are promising: on synthetic data, we show that our proposed method does indeed diversify among time series with highly similar structural breaks and enjoys advantages over existing metrics between sets. On real data, experiments illustrate that our proposed optimization method performs well relative to nine other commonly used options, producing the second-highest returns, the lowest volatility, and second-lowest drawdown. The main implication for this method in portfolio management is reducing simultaneous asset shocks and potentially sharp associated drawdowns during periods of highly similar structural breaks, such as a market crisis. Our method adds to a considerable literature of portfolio optimization techniques in econometrics and could complement these via portfolio averaging.
}


\begin{document}

\section{Introduction}
\label{sec:intro}

One of the oldest and most important tasks in the field of econometrics is the analysis, forecasting, and optimization of financial risk. This may be conducted at the level of an individual stock, an entire sector, or a judiciously chosen portfolio. The most common measure of risk is portfolio variance, popularized by \cite{Markowitz1952} in his seminal work. Markowitz' mathematical derivations assumed the Gaussianity of financial returns. Subsequently, returns of financial assets were shown to be non-Gaussian and fat-tailed in several works \citep{Mandelbrot1963,Fama1965}, prompting analysts to seek alternative measures of risk. Notably, tail risk measures such as value-at-risk \citep{Duffie1997,Braione2016,Khraibani2018} or conditional value-at-risk/expected shortfall \citep{Krause2014,tsay_analysis_2010,long_statistical_2020,ullah_conditional_2022} have proven useful to guard against the greatest possible losses amid a financial crisis. In this paper, we propose an alternative approach to and measure of risk reduction, especially during a crisis, focusing on the diversification of assets away from simultaneous asset shocks, specifically in the form of coincident structural breaks.

Modern portfolio theory provides a framework for determining an allocation of weights in an investment portfolio by optimizing a specific objective function. The idea was first introduced by \cite{Markowitz1952} and has progressed considerably since then. Markowitz' fundamental contribution was the concept of diversification among stock portfolios, rather than analyzing risk and return on an individual security basis, one of the most seminal breakthroughs in econometrics. One of the most notable advancements was the work of \cite{Sharpe1966}, who proposed a measure of risk-adjusted returns in financial portfolios, the Sharpe ratio. This ratio is an indication of the potential reward in any candidate investment relative to its risk. The standard mathematical representation of the Sharpe ratio is the following optimization problem: given a collection of $n$ assets, let $R_i$ be the historical returns for the $i$th asset in a collection, $\Sigma$ be the matrix of historical covariances between stocks, $R_{f}$ the risk-free rate, and $w_i$ the weights of the portfolio. One maximizes the Sharpe ratio, which we define by the following optimization problem:
\begin{align}
\label{eq:Sharpeobjectionfn}
\text{Maximize: } \frac{\sum^{n}_{i=1} w_{i} R_{i} - R_f}{ \sqrt{\boldsymbol{w}^{T} \Sigma \boldsymbol{w}}  }, \\
\text{subject to: } 0 \leq w_{i} \leq 1, i = 1,...,n, \label{constrant1} \\
\sum^{n}_{i=1} w_{i} = 1.\label{constrant2}
\end{align}
This objective function (\ref{eq:Sharpeobjectionfn}) selects an allocation of weights based on a trade-off between portfolio returns and variance. Returns are estimated from historical returns as $\E(R_{p}) = \sum^{n}_{i=1} w_{i} R_{i}$, while variance is estimated via $\sigma_{p}^{2} = \boldsymbol{w}^{T} \Sigma \boldsymbol{w}$.

One may also impose certain conditions, depending on the context, which manifest as constraints, to accompany the objective function. The most common constraints, which we impose above and throughout the paper, are $0 \leq w_{i} \leq 1, i = 1,...,n$ (\ref{constrant1}) and $\sum^{n}_{i=1} w_{i} = 1$ (\ref{constrant2}). We also assume $R_f=0$ throughout. These conditions require all portfolio assets to be invested and prohibit short selling, respectively. Weights are chosen by maximizing the objective function subject to such conditions. There is a wealth of other conditions that may be imposed, which is discussed further in the paper. 

\subsection{Overview of portfolio optimization}

There has been significant research within the applied mathematics, computer science, and econometrics communities building upon Markowitz's mean-variance model \citep{Markowitz1952,Sharpe1966}. A variety of portfolio optimization frameworks have explored alternative objective functions utilizing risk measures other than standard volatility \citep{Almahdi2017,Calvo2014,Soleimani2009,Vercher2007,Bongini2002}. Many authors in the field have taken existing theory and methodologies for the problem of portfolio selection and optimization \citep{Bhansali2007,Moody2001,MagdonIsmail}, including statistical mechanics \citep{Zhao2016,Li2021_portfolio}, clustering \citep{Iorio2018,Len2017}, fuzzy sets \citep{Tanaka2000,Ammar2003,Kocadal2015}, graph theory \citep{james_georg}, regularization \citep{Fastrich2014,Li2015,Pun2019} regression trees \citep{Cappelli2021}, and multiobjective optimization \citep{Mansour2019,Lam2021}. For further details, a review of such techniques for portfolio optimization was conducted by \cite{Milhomem2020}. Most importantly, for this paper, we must especially acknowledge the influence of statistical physics and econophysics. Many modern analyses of traditional financial markets \citep{Laloux1999,Eisler2006,james_georg,Alves2020}, cryptocurrency markets \citep{Matjaz_crypto,Drod2020,Drod2020_entropy,Drod2018,Drod2019,Wtorek2020,Sigaki2019}, and portfolio optimization \citep{james2022_stagflation,james_arjun} have built upon methods developed in econometrics or inspired by physics \citep{Mantegna2000,Podobnik2009,james2021_mobility,Wang2006,Liu1999,Gopikrishnan1998,james2021_MJW,Basalto2007,Basalto2008,james2021_crypto2,Dose2005,Valenti2018,Fister2021,Wang2020_Matjaz,james2021_TVO,james2021_hydrogen,Drod2021_entropy,james2021_olympics,James2021_geodesicWasserstein,james2023_hydrogen2,james2020_Lp,james2022_guns,james2022_CO2,James2023_terrorist}.

In particular, there has been a wealth of work that has modified the Sharpe ratio to penalize downside risk specifically. \cite{Sortino1991} were pioneers in this, modifying the Sharpe ratio directly to only penalize downside variance. Since then, various frameworks have been developed to directly target loss reduction, including value at risk models \citep{Campbell2001,Alexander2002} and the mean--semivariance framework \citep{Ballestero2005,Boasson2011,BenSalah2018}.

Finally, a substantial body of work has examined the difficulties provided by more complex investment constraints. \cite{Jin2016} provided a review of typical constraints in an asset allocation problem, as well as advances in algorithmic procedures \citep{Meghwani2017,Liagkouras2015,Liagkouras2017,Lwin2014}. In particular, cardinality constraints \citep{Anagnostopoulos2011} yield nonconvex sets (unions of lower-dimensional simplices) over which to perform optimization, providing a challenge to standard methods of convex optimization and producing NP-hard problems \mbox{\citep{Shaw2008}.}

\textls[-15]{In our paper, we compare our methodology primarily with those defined over long periods of training data. The primary reason we do this is to provide the algorithm adequate time to learn the} systemic similarity in various asset's change point propagation. Given that change points indicate major shifts in underlying return dynamics, one can appreciate that these changes do not occur frequently. In fact, asset classes with especially low levels of beta (market risk) may only produce change points in the most extreme market conditions. Given that we wish to compute distances between change point propagation, which should include a sufficient number of change points, it is necessary that the training period is of significant length. That being said, our algorithmic approach may not need to be updated (and require the model being retrained) as frequently as other methods---as we are concerned with optimizing over very low frequency signals, which are unlikely to change~quickly.

\subsection{Overview of change point detection methods}

Many domains in the physical and social sciences are interested in the identification of structural breaks in various data sets. \cite{Ranshous2015} and \cite{Akoglu2014} recently provided an overview of anomaly detection methods within the context of network analysis, which can be used to identify relations among entities in high-dimensional data. \cite{Koutra2016} determined change points (structural breaks) in dynamic networks via graph-based similarity measures, while \cite{James2021_crypto} analyzed change points in cryptocurrencies.

In the more econometric and statistical literature, focused on time series data, researchers have developed change point models driven by hypothesis tests, where $p$-values allow scientists to quantify the confidence in their algorithm \citep{Moreno2013,Bridges2015,Peel2015}. Change point algorithms generally fall within statistical inference (namely Bayesian) or hypothesis testing frameworks. Bayesian change point algorithms \citep{Barry1993,Xuan2007,Adams2007,james2021_spectral} identify change points in a probabilistic manner and allow for subjectivity through the use of prior distributions, but they suffer from hyperparameter sensitivity and do not provide statistical error bounds ($p$-values), often leading to a lack of~reliability. 

Within hypothesis testing, \cite{RossCPM} outlined algorithmic developments in various change point models initially proposed by \cite{Hawkins1977}. Some of the more important developments in recent years include the work of \cite{Hawkins2003} and  \cite{Ross2012,Ross2013,Ross2014}. \cite{RossCPM} recently created the CPM package, which allows for flexible implementation of various change point models on time series data. Given the package's ease of use, flexibility, and efficient implementation, we build our methodology on this suite of algorithms.

\subsection{Overview of semi-metrics} 

The application of metric spaces has provided the groundwork for research advancement in various areas of machine learning. In addition to more traditional metrics, such as the Hausdorff and Wasserstein, semi-metrics, which may not satisfy the triangle inequality property of a metric, have been used successfully in various machine learning applications. An overview of such (semi-)metrics and applications was recently provided by \cite{Conci2017}. The three primary applications include image analysis \citep{Baddeley1992,Dubuisson1994,Gardner2014}, distance between fuzzy sets \citep{Brass2002,Fujita2013,Gardner2014,Rosenfeld1985}, and computational methods \citep{Eiter1997,Atallah1983,Atallah1991,Shonkwiler1989}. 
 
 More recently, a review and computational analysis of various (semi-)metrics was undertaken by \cite{James2020_nsm} in measuring distance between time series' sets of structural breaks.

\subsection{Motivation and structure of this paper}
\label{sec:motivation}

This paper aims to draw upon the aforementioned fields to yield a new approach to portfolio optimization with several benefits. While other existing methods aim to reduce downside risk directly \citep{Boasson2011,BenSalah2018}, we consider the significant shifts in asset behavior around structural breaks as a kind of ``root cause'' for simultaneous drawdown and paramount to avoid as the highest priority. Thus, we introduce the framework of using distances between structural breaks in our objective function; this necessitates distance measures between finite sets. In addition, we claim novelty in the precise methodology to measure discrepancy between finite sets. While previous research typically uses the Hausdorff or Wasserstein \citep{Basalto2007,Basalto2008} metric between sets, we use our own semi-metric, with favorable theoretical and empirical properties. Thus, our primary contributions in this paper are a novel framework of using structural breaks in portfolio optimization and a ``proof of concept'' via our specific implementation, as well as its validation. Validation is performed via simulated and real data, as well as a sequence of new propositions contrasting our chosen distance with previous options.

Our motivation in this paper is both theoretical and experimental. Theoretically, we investigate the use of distances between structural breaks in a broad attempt to highlight their potential utility in portfolio optimization applications, which we are unaware exists in the literature. We prove numerous properties of our particular choice of discrepancy between finite sets relative to existing alternatives. Additionally, experimentally, we offer a particular technique of portfolio optimization with promising results relative to existing options. The contribution of this paper goes beyond the specific methodology proposed; utilizing structural breaks may have numerous research directions. First, our specific methodology could be used within an ensemble framework, where our model could be one of several portfolio optimization procedures used to determine optimal portfolio weights (akin to model stacking and other ensemble-based methods). Second, our work could encourage researchers outside of econometrics and investment management specifically, but there is a general interest in allocating weights to various features, where each feature exhibits some sort of penalty in a time-varying fashion. This would be of particular interest in settings where the panel of data is especially stationary.

Thus, our paper is structured as follows. In Section \ref{sec:framework}, we outline our framework for optimization using structural breaks, including a detailed explanation of the general benefits of the framework. In Section \ref{sec:theoreticalproperties}, we explain the specific theoretical properties of our precise semi-metric compared with the existing Hausdorff or Wasserstein options to measure discrepancy between finite sets to explain the benefits of our precise choice. In Section~\ref{sec:synthetic1}, we use synthetic time series to show these benefits over other metrics more concretely in illustrated examples. Section \ref{sec:synthetic2} then performs a sample allocation of capital within a typical constrained optimization problem, featuring constraints frequently required by financial practitioners. In Section \ref{sec:realdata}, we apply our method to real data across judiciously chosen training and testing  periods, where we compare our methodology against nine others from the optimization literature. We conclude in Sections \ref{sec:conclusion} and \ref{sec:newconclusion}, summarizing the utility of our framework in the context of a typical asset allocation scenario.

\section{Proposed semi-metric change point optimization framework} 
\label{sec:framework}

Our main contribution is to adapt the classical penalty function involving variance with one related to structural breaks. In many circumstances, variance is a suitable measure in a financial securities context. However, it is not without its limitations, and there are several reasons why a penalty function related to structural breaks may be a suitable alternative or complement to the covariance measure between two time series:
\begin{enumerate}

    \item Covariance is computed as an expectation $\Cov(X,Y)=\E(X-\E X)(Y-\E Y)$, which is an average (integral) over an entire probability space. In a financial context, this computes an average over time; in modern financial markets, especially since the global financial crisis, most time periods are bull markets, with most assets performing quite well together. As such, assets that rise together in a bull market but actually exhibit distinct dynamics may be erroneously identified as similar.
    
    \item Covariance fails to capture dissimilarity between time series during periods of market crisis and erratic behavior. Investors are often particularly concerned with the robustness of their portfolio during such times. Portfolios that are optimized using covariance as a risk measure fail to determine the impact of various asset combinations during times of market crisis. For instance, if two assets are simultaneously acting erratically, they may actually be negatively correlated during this time. If they are both included in a portfolio, this would increase rather than reduce erratic behavior. Structural breaks herald erratic behavior, so using distances between breaks in the objective function may better separate out erratic behavior in a portfolio. 
    
    \item Investors are also interested in peak-to-trough measures of asset performance, that is, the size of a drop in returns from a local maximum to a local minimum. Optimization algorithms using covariance measures fail to identify and minimize peak-to-trough behavior. However, distances between sets of structural breaks (in the mean, variance, and other stochastic quantities) are better equipped to identify how similar two time series are with respect to peak-to-trough measures. Thus, they may suitably allocate weights to minimize these precipitous drops. 
    
    \item While various methods of portfolio optimization target downside risk directly, we believe that structural breaks may be a kind of ``root cause'' of the greatest erratic behavior and simultaneous downside risk, and thus are of the greatest priority to diversify away from.
    
\end{enumerate}

We formulate our new objective function to penalize structural breaks and their associated erratic behavior. We use the MJ$_p$ family of semi-metrics of \cite{James2020_nsm}. Given $p>0$ and two nonempty finite sets $A,B \subset \mathbb{R}$ (or an arbitrary metric space), this is defined as
\begin{equation}
\label{eq:MJdefn}
    d^p_{MJ}({A},{B}) = \Bigg(\frac{\sum_{b\in B} d(b,A)^p}{2|B|} + \frac{\sum_{{a} \in {A}} d(a,B)^p}{2|A|} \Bigg)^{\frac{1}{p}}.
\end{equation}
where $d(a,B)$ is the minimal distance from a point $a \in A$ to the finite set $B$. We note $d^p_{MJ}(A,B)=0$ if and only if $A,B$. As discussed by \cite{James2020_nsm}, varying $p>0$ produces a family of semi-metrics, where larger values of $p$ exhibit greater adherence to the triangle inequality property, but worse sensitivity to outliers. In our implementation, we select $p=0.5$ due to its good performance with outlier sensitivity and the strong possibility of outliers in this context. Indeed, it is likely that some assets are impacted by market dynamics to which others are immune, which will yield outlier assets. We discuss distances between finite sets, their properties, and how we arrive at our family of semi-metrics in \ref{Appendix_distances}.

We compute a distance matrix $D_{ij}$ as follows: following a suitable change point algorithm (ours is described in Appendix \ref{Appendix_CPD}), let asset $i$ have set of structural breaks $S_i,i=1,...,n$. Then, we form
\begin{align}
\label{eq:MJchoice}
    D_{ij}=d^{0.5}_{MJ}(S_i,S_j).
\end{align}
In this paper, every set of structural breaks, simulated or real, is nonempty, so this computation is possible. Next, we transform our distance matrix into an affinity matrix, which mimics the properties of a covariance matrix:
\begin{equation}
\label{eq:affinity}
    A_{ij} =  1 - \frac{D_{ij}}{\max D},  \forall i,j.
\end{equation}
 Two 
 assets have correlation equal to $1$ if and only if they are perfectly correlated; analogously, $A_{ij}=1 $ if and only if  $d(S_i,S_j)=0$, meaning the two assets have identical structural~breaks. 

\begin{Remark}
\label{remark:outliers}
We note that the denominator $\max D$ may be influenced by a single outlier time series. Thus, it is particularly important to choose our semi-metric (particularly the value of $p$) to handle outlier elements with care. We justify the benefits of choosing a smaller value of $p$, in this case $p=\frac12$, in Proposition \ref{prop:Hausdorff}, Corollary \ref{cor:Hausdorff}, and Appendix \ref{Appendix_distances}.
\end{Remark}

In the context of Markowitz portfolio optimization, weights are chosen to maximize return while reducing total variance; this introduces more stocks with lower correlation, increases diversification, and reduces systematic risk in the portfolio. We modify this insight, allocating weights that maximize return while reducing affinity between sets of structural breaks, hence maximizing the spread between erratic behavior. To do so, we substitute our adjusted affinity matrix $A$ for the original covariance matrix $\Sigma$ and optimize a new risk-adjusted return measure with respect to portfolio weights. We term this the MJ ratio objective function (\ref{eq:MJ}), which we define in the following optimization problem:
\begin{align}
    \label{eq:MJ}
\text{Maximize: }\frac{\sum^{n}_{i=1} w_{i} R_{i} - R_{f}}{\sqrt{\boldsymbol{w}^{T} A \boldsymbol{w}}}, \\
\text{subject to: } 0 \leq w_{i} \leq 1, i = 1,...,n, \label{MJratioconstrant1} \\
\sum^{n}_{i=1} w_{i} = 1.\label{MJratioconstrant2}
\end{align}
Essentially, 
 this method retains the estimation of returns exactly as in the Sharpe ratio, $\E (R_{p}) = \sum^{n}_{i=1} w_{i} R_{i}$ and substitutes variance $\sigma_{p}^{2} = \boldsymbol{w}^{T} \Sigma \boldsymbol{w}$ with a new denominator $\Omega_{p}^2 = \boldsymbol{w}^{T} A \boldsymbol{w}$, whose purpose is to ``spread out'' various assets' structural breaks.

Throughout the paper, we always retain at least the same constraints as Section \ref{sec:intro}, $0 \leq w_{i} \leq 1, i = 1,...,n$ (\ref{MJratioconstrant1}) and $\sum^{n}_{i=1} w_{i} = 1$ (\ref{MJratioconstrant2}). In subsequent sections, we also impose additional real-world constraints, such as upper and lower bounds on weights, and discuss how our method would work with other constraints frequently used in investment policy statements. Our method is flexible enough to vary such constraints, with no increase in complexity, provided such constraints result in a convex set of optimization. For more difficult constraints such as cardinality and preassignment constraints, our method could be combined with advances in the literature for efficient optimization over such spaces \citep{Liagkouras2017}.

\begin{Remark}
One could also use this approach (substituting variance for affinity between structural breaks) to modify alternative existing portfolio selection methods, such as minimum variance optimization. Traditionally, this is defined as the following optimization problem:
\begin{align}
    \text{Minimize: } \boldsymbol{w}^{T} \Sigma \boldsymbol{w}, \\
    \text{subject to } \sum^{n}_{i=1} w_{i} R_{i}=P, \label{MVconstraint0} \\
    0 \leq w_{i} \leq 1, i = 1,...,n, \label{MVconstrant1} \\
\sum^{n}_{i=1} w_{i} = 1.\label{MVconstrant2}
\end{align}
Typically, the desired level of returns $P$ is selected according to the risk appetite of the investor. One can also incorporate a risk-free asset by including it as one of the permissible asset classes, with a term $R_0=R_f$.

One could then write an equivalent of minimum variance optimization in our new context of structural breaks by formulating an optimization problem as follows:
\begin{align}
    \text{Minimize: } \boldsymbol{w}^{T} A \boldsymbol{w}, \\
    \text{subject to } \sum^{n}_{i=1} w_{i} R_{i}=P,  \label{MJnewconstraint0} \\
    0 \leq w_{i} \leq 1, i = 1,...,n, \label{MJnewconstrant1} \\
\sum^{n}_{i=1} w_{i} = 1.\label{MJnewconstrant2}
\end{align}
Alternatively, one could formulate an analogy to the global minimum variance problem. This would be formulated simply by removing condition (\ref{MJnewconstraint0}) as follows:
\begin{align}
    \text{Minimize: } \boldsymbol{w}^{T} A \boldsymbol{w}, \\
     \text{subject to } 0 \leq w_{i} \leq 1, i = 1,...,n, \label{MJnewconstrant1_1} \\
\sum^{n}_{i=1} w_{i} = 1.\label{MJnewconstrant2_1}
\end{align}
\end{Remark}

\section{Theoretical properties}
\label{sec:theoreticalproperties}
In this section, we examine the mathematical properties of our proposed objective function and procedure and explain our choice of distance function between sets, including an analysis of alternatives.

\begin{Proposition}
\label{prop:compact}
The MJ ratio, as presented in (\ref{eq:MJ}), can be maximized on the chosen domain of weights, and the maximum can be determined analytically.
\end{Proposition}
\begin{proof}
First, we note that the matrix $A$ is not necessarily positive semidefinite, so standard arguments regarding the optimization of the Sharpe ratio do not apply \textit{mutatis muntandis} to the MJ ratio. Instead, we require a continuity and compactness argument. Due to the conditions on the weights, the ratio is optimized over a space $S=\{ w_i: 0 \leq w_i \leq 1, \sum_{i=1}^n w_i = 1 \}$. This is a compact space, specifically a $(n-1)-$simplex. By the definition of (\ref{eq:affinity}), all entries of $A$ are non-negative, with diagonal elements equal to 1. Thus, $w^T A w$ is a continuous function on $S$ that attains only positive values, and so the denominator of (\ref{eq:MJ}) is positive on the whole space $S$. This implies the MJ ratio is a well-defined continuous function on $S$. Since $S$ is compact, it must achieve a global maximum on $S$.

Finally, since $S$ is a $(n-1)-$simplex, one can examine and test the critical points within the simplex and use Lagrange multipliers on the boundary to find all possible maxima and test them. In our implementation, we determine the optimal weights with a simple grid search.
\end{proof}

\begin{Proposition}[Method complexity]
Suppose we have $n$ assets indexed $i=1,...,n$ over a time period of length $T$. Assume the weights have only the minimal constraints $0 \leq w_i \leq 1, \sum_{i=1}^n w_i = 1$. Then, the computational cost of the weight selection methodology described in Section \ref{sec:framework} is $O(n^2T + nT^2)$.
\end{Proposition}

\begin{proof}
As explained in \ref{Appendix_CPD}, the selection of change points for a single asset has a running time of $O(T^2)$ due to the two-phase procedure. Thus, the selection of change points $S_1,...,S_n$ for all assets has cost $O(nT^2)$. For each $i,j$, the computation of $d^p_{MJ}(S_i,S_j)$ involves at most $O(|S_i|+|S_j|)$ comparisons between elements. Let $m=\max_i |S_i|$. This means the computation of each pairwise distance $d^p_{MJ}(S_i,S_j)$ is of complexity $O(m)$. Thus, the construction of the full distance matrix $D$ and associated affinity matrix $A$ is of complexity $O(mn^2)$. To select the weights, only the historical returns $R_i$ and the matrix $A$ are needed. We use sequential quadratic programming, which has a complexity cost of $O(n^2)$ when performed over a convex set with a fixed tolerance bound.

Overall, the total cost of our procedure is $O(nT^2 + mn^2 + n^2)$; the three steps are each implemented in C\texttt{++}. We can use the simple bound $m\leq T$ to gain the final bound $O(n^2T + nT^2)$.
\end{proof}

\begin{Remark}
In our implementation, we consistently train our algorithm over long periods to appropriately learn relationships in structural breaks. Thus, it is usually the case that $n << T$. For example, a typical portfolio manager would have at most $n=1000$ assets to choose from, while we train over $T=2051$ days. Thus, the computational cost simplifies to $O(nT^2)$, with the $T^2$ operations implemented efficiently in C\texttt{++}. Thus, we can deduce that our method scales well with large numbers of stocks to choose from, with just a linear increase in complexity with the number of~assets.

Furthermore, our complexity is unchanged with real-world upper and lower bounds on the weights $c_i \leq w_i \leq C_i$. These constraints, common in investment policy statements, still produce a convex and compact set over which we select the weights $w_i$, so the complexity is unchanged. As discussed in Section \ref{sec:realdata}, these are the only additional constraints we impose in our experimentation, a common feature of real-world policy statements \citep{RussellPolicy}. When additional constraints are imposed, our optimization domain may be nonconvex. However, it will always be compact, so Proposition~\ref{prop:compact} holds. Efficient optimization of our objective function would require a combination with recent work in optimizing over domains with numerous nonconvex constraints imposed \citep{Liagkouras2015,Liagkouras2017,Lwin2014,Meghwani2017,Shaw2008}.

\end{Remark}

In the following two propositions, we justify our selection of distance measure between finite sets, specifically two advantages it has over the popular Hausdorff and Wasserstein metrics between sets.

\begin{Definition}[Hausdorff metric]
\label{def:Hausdorff}
Let $S,T$ be closed bounded subsets of $\mathbb{R}$ (or an arbitrary metric space). Their Hausdorff distance is defined by
\begin{align}
    d_{H}(S,T) =&  \max \left( \sup_{s \in S} d(s,T), \sup_{t \in T} d(t,S) \right), \\
    =&  \sup \{ d(s,T), s \in S; d(t,S), t \in T \},
\end{align}
where $d(s,T)=\inf_{t \in T} d(s,t)$ is the infimum distance from $s$ to $T$.
\end{Definition}
One could conceivably use this metric to measure distance between structural breaks, rather than our semi-metric. However, the Hausdorff metric suffers from substantial sensitivity to outliers, as observed by \cite{Baddeley1992}. We formalize this in the following proposition:
\begin{Proposition}
\label{prop:Hausdorff}
Let $T=\{t_1,...,t_n\}$ and $S$ be fixed. Fixing all but one element, if $t_n \to \infty$ acts as an outlier, then the asymptotic behavior of the Hausdorff and MJ$_p$ distances are as follows:
\begin{align}
\label{eq:firstasymptotic}
d_H(S,T) \sim |t_n|, \hspace{8.5mm}  d^{p}_{MJ}(S,T) \sim \frac{|t_n|}{(2|T|)^{\frac{1}{p}}}, \\ \text{i.e.} \hspace{2mm} \lim_{|t_n| \to \infty} \frac{d_H(S,T)}{|t_n|} = 1,  \lim_{|t_n| \to \infty} \frac{d^{p}_{MJ}(S,T)}{|t_n|} = \frac{1}{(2|T|)^{\frac{1}{p}}}.
\label{eq:secondasymptotic}
\end{align}
\end{Proposition}
\begin{proof}
For both $d_H(S,T)$ and $d^{p}_{MJ}(S,T)$, the only term that increases as $|t_n| \to \infty$ is $d(t_n,S)$, which increases asymptotically with $|t_n|$. The result follows immediately for $d_H$ and follows for $d^{p}_{MJ}$ by inspecting the coefficient $\frac{1}{2|T|}$ that accompanies the term $d(t_n,S)$.
\end{proof}

\begin{Corollary}
\label{cor:Hausdorff}
Let $0<p<q$ and adopt the assumptions of Proposition \ref{prop:Hausdorff}. Then, $d_H(S,T)$ exhibits worse asymptotic outlier sensitivity than $d^q_{MJ}$, which itself is worse than $d^p_{MJ}$.
\end{Corollary}
\begin{proof}
If $p<q$, then $1<(2|T|)^p < (2|T|)^q$. It follows that
\begin{align}
    \frac{1}{(2|T|)^\frac{1}{p}} < \frac{1}{(2|T|)^\frac{1}{q}} < 1.
\end{align}
Thus, 
 the asymptotic coefficient of $|t_n|$ is the least for $d^p_{MJ}$, then $d^q_{MJ}$, and then $d_H$, as shown in (\ref{eq:firstasymptotic}) and (\ref{eq:secondasymptotic}). That is, a single element has the least influence on the increasing values of $d^p_{MJ}$ than on $d^q_{MJ}$, and less so than $d_{H}$.

\end{proof}

As a consequence of this outlier sensitivity property, the Hausdorff metric $d_H$ may grant an excessively high distance, and hence a low affinity, based on a single outlier structural break. In particular, two time series that have quite similar structural breaks (and hence erratic behavior profiles) may be granted low affinity and both be included in a portfolio based on just one structural break. On the other hand, the MJ$_p$ semi-metric handles outliers well, and increasingly well with small values $p$, such as our choice in implementation $p=0.5$. We illustrate an example of this in Section \ref{sec:synthetic1} and explain this further in Appendix \ref{Appendix_distances}.

The other metric occasionally used to measure distance between finite sets is the Wasserstein metric. To be precise, it is most frequently employed between probability measures on a metric space, such as $\mathbb{R}$, as follows: let $\mu,\nu$ be probability measures on $\mathbb{R}$, and $q \geq 1$, then
\begin{align}
\label{eq:Wasserstein}
    W_{q} (\mu,\nu) = \inf_{\gamma} \bigg( \int_{\mathbb{R} \times \mathbb{R}} |x-y|^q d\gamma  \bigg)^{\frac{1}{q}}.
\end{align}
This infimum is taken over all joint probability measures $\gamma$ on $\mathbb{R}\times \mathbb{R}$ with marginal probability measures $\mu$ and $\nu$. The formula (\ref{eq:Wasserstein}) is difficult to compute in general, but in the case where $\mu,\nu$ have cumulative distribution functions $F,G$ on $\mathbb{R}$, there is a simple representation \citep{DelBarrio}:
\begin{align}
\label{eq:computeWasserstein}
  W_{q} (\mu,\nu) =  \left(\int_{0}^1 |F^{-1} - G^{-1}|^q dx\right)^\frac{1}{q},
\end{align}
where $F^{-1}$ is the inverse cumulative distribution function, or more precisely, quantile function, associated to $F$ \citep{Gilchrist2000}. One can then use this to define a metric between finite sets $S,T$. One associates to each set a probability measure defined as a weighted sum of Dirac delta measures
\begin{align}
\label{eq:Wasserstein delta}
    \mu_S=\frac{1}{|S|}\sum_{s \in S}\delta_s.
\end{align}
Then, the Wasserstein metric between sets $S,T$ can be defined as $d_W^q(S,T):=W_q(\mu_S,\mu_T)$ and computed with (\ref{eq:computeWasserstein}). One could conceivably use this metric to measure distance between structural breaks instead of the Hausdorff metric. However, the Wasserstein metric has a property that makes it unsuitable in our context. Using the definition (\ref{eq:Wasserstein delta}) and the Equation (\ref{eq:computeWasserstein}), the Wasserstein metric has a geometric property with respect to translation, $d_W^q(S, S+a)=|a|$. However, this is unsuitable for measuring the distance between sets with high intersection. We formalize these remarks in the following proposition:
\begin{Proposition}
\label{prop:Wasserstein}
If $|S\cap T|=r,$ the following inequality holds: 
\begin{align}
\label{eq:intersectionineq}
d^p_{MJ}(S,T) \leq \left[1-\frac{r}{2}\Big(\frac{1}{|S|}+ \frac{1}{|T|}\Big)\right]^{\frac{1}{p}}d_H(S,T).   
\end{align}
No such inequality holds for Wasserstein metric. Given a set $S$ and its translation $S+a$ for some $a\in \mathbb{R}$, the Wasserstein metric has the property that $d_W^q(S, S+a)=|a|$. As a consequence, even with $|S\cap T|=|S|-1=|T|-1$, it is possible for $d_W^q(S,T)$ to coincide with $d_H(S,T)$.
\end{Proposition}
\begin{proof}
Examining the definition (\ref{eq:MJdefn}), any $d(s,T)$ or $d(t,S)$ term with $s \in S\cap T $ or $t \in S\cap T$, respectively, vanishes. Any other $d(s,T),d(t,S)$ term is at most $d_H(S,T)$. So,
\begin{align}
    d^p_{MJ}({S},{T}) \leq \left[ \frac{1}{2|S|}(|S|-r) + \frac{1}{2|T|}(|T|-r)\right] d_H(S,T),
\end{align}
which gives the inequality after simplifying. Turning to the Wasserstein metric, let $S=\{s_1,...,s_n\} \subset \mathbb{R}$ be a set with $s_1<s_2<...<s_n$ and $a \in \mathbb{R}$ a translate. Then, $S+a=\{s_1+a,...,s_n+a\}$. By  (\ref{eq:computeWasserstein}) and (\ref{eq:Wasserstein delta}), $d_W^q(S, S+a)$ can be computed as 
\begin{align}
\label{eq:nextthing}
    \left(\int_{0}^1 |F^{-1} - G^{-1}|^q dx\right)^\frac{1}{q},
\end{align}
where $F^{-1},G^{-1}$ are the quantile functions associated to $\mu_S$ and $\mu_{S+a}$. By integrating $\mu_S$ and $\mu_{S+a}$, we can see that $F,G$ are piecewise constant increasing step functions:
\begin{align}
    F=\sum_{j=1}^{n-1} \frac{j}{n} \mathbbm{1}_{[s_j,s_{j+1})} + \mathbbm{1}_{[s_n,\infty)}, G= \sum_{j=1}^{n-1} \frac{j}{n}\mathbbm{1}_{[s_j+a,s_{j+1}+a)} + \mathbbm{1}_{[s_n+a,\infty)}.
\end{align}
It follows that their respective quantile functions are determined almost everywhere as
\begin{align}
    F^{-1}=\sum_{j=1}^{n} s_j \mathbbm{1}_{(\frac{j-1}{n},  \frac{j}{n})}, G^{-1}=\sum_{j=1}^{n} (s_j+a) \mathbbm{1}_{(\frac{j-1}{n},  \frac{j}{n})}.
\end{align}
It follows quickly that $G^{-1} - F^{-1}$ is simply a constant function on $(0,1)$ with value $a$, so the expression (\ref{eq:nextthing}) simplifies to $|a|$. This concludes the second statement of the proposition.

Finally, let $S=\{0,1,...,n-1\}$ and $T=\{1,2,...,n\}$. As $T=S+1$ is a translate of $S$, we have shown that $d_W^q(S,T)=1=d_H(S,T)$. However, $|T|=|S|=n$, while $|S \cap T|=n-1$, showing that no such inequality as (\ref{eq:intersectionineq}) holds for the Wasserstein metric.
\end{proof}

As a consequence of this translation property, the Wasserstein metric $d_W^q$ may excessively grant an excessively high distance, and hence a low affinity, to two sets with a very high intersection. For example, if two sets of structural breaks are $A=\{100,200,...,900\}$ and $B=\{200,300,...,1000\}$, then $d^p_{MJ}$ will reflect the high intersection and similarity between the sets $A$ and $B$, while $d_W^q$ will not. Indeed, for this example, $d_W^q(A,B)=100$, while $d^p_{MJ}(A,B)=100\big(\frac{1}{9}\big)^{\frac{1}{p}}$. That is, only the latter semi-metric assigns these remarkably similar sets of structural breaks with a low distance, hence a high affinity.

The Wasserstein metric would grant two time series that have quite similar structural breaks (and hence erratic behavior profiles) low affinity, and hence could include both in a portfolio. This would be a mistake, as such structural breaks as given by $A$ and $B$ in fact have almost all elements in common and should be assigned high affinity, so the portfolio will not choose them both. We illustrate an example of this in Section \ref{sec:synthetic1}.

\section{Simulation study}
In this section, we perform two experiments involving simulated time series with specified structural breaks. The first experiment illustrates the computation of the similarity between sets of structural breaks, comparing our MJ$_p$ distance with the Hausdorff and Wasserstein metrics. The primary purpose of this experiment is to illustrate the benefits of our chosen discrepancy measure over alternative and existing metrics. Our examples are chosen to exemplify, via a small number of time series, how the properties proven in Section \ref{sec:theoreticalproperties} matter for real data. The second experiment illustrates the allocation of assets in a sample optimization problem, together with constraints  typical of an investment policy~statement.

\subsection{Synthetic data simulation}
\label{sec:synthetic1}
\begin{figure}[H]
    \begin{subfigure}[b]{0.49\textwidth}
        \includegraphics[width=\textwidth]{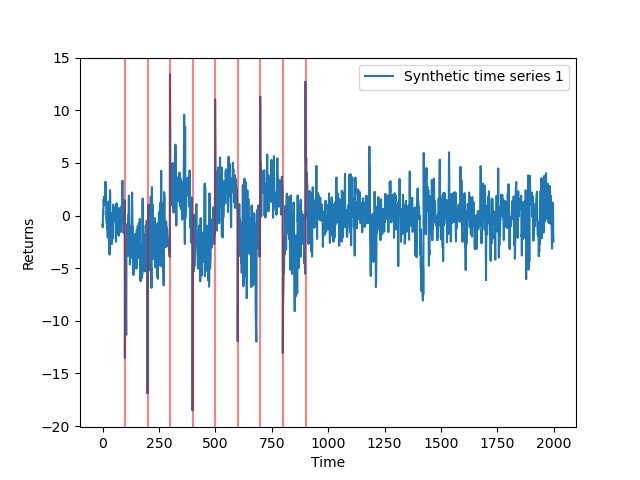}
        \caption{}
        \label{fig:Synthetic_ts_1}
    \end{subfigure}
    \begin{subfigure}[b]{0.49\textwidth}
        \includegraphics[width=\textwidth]{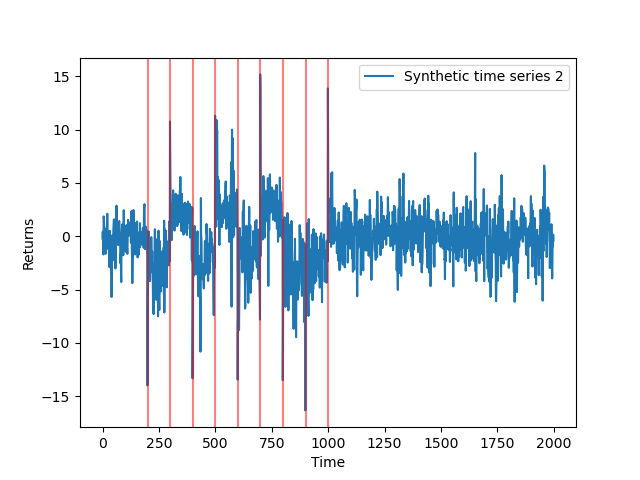}
        \caption{}
        \label{fig:Synthetic_ts_2}
    \end{subfigure}\\\vspace{-2pt}
    
        \begin{subfigure}[b]{0.49\textwidth}
        \includegraphics[width=\textwidth]{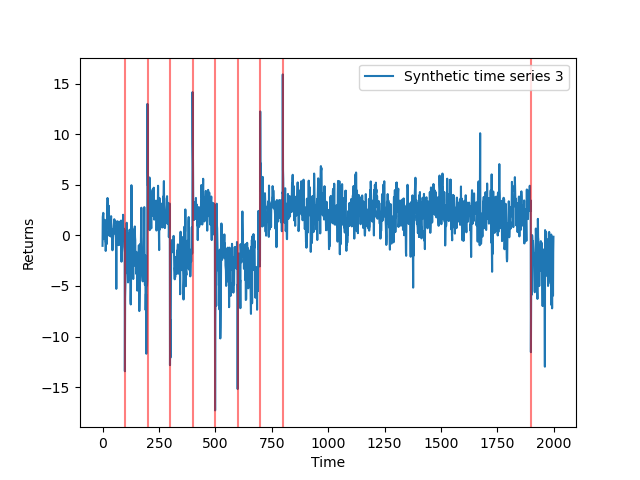}
        \caption{}
        \label{fig:Synthetic_ts_3}
    \end{subfigure}
        \begin{subfigure}[b]{0.49\textwidth}
        \includegraphics[width=\textwidth]{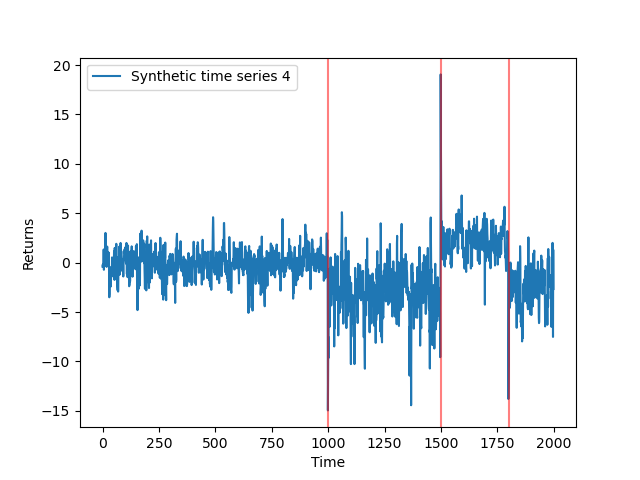}
        \caption{}
        \label{fig:Synthetic_ts_4}
    \end{subfigure}
    \caption{\textls[-15]{Four synthetic time series (\textbf{a}--\textbf{d}) exhibiting dependence and correlation in jump behavior. The red annotated lines represent specified structural breaks. The structural breaks are chosen carefully to relate to Corollary \ref{cor:Hausdorff} and Proposition \ref{prop:Wasserstein}, where we discuss some undesirable properties of the Hausdorff and Wasserstein metrics, respectively. (\textbf{a},\textbf{b}) are chosen to have all but one break in common; the Wasserstein metric allocates an excessive distance. (\textbf{c}) is chosen to feature a} single outlier break compared with (\textbf{a},\textbf{b}); the Hausdorff metric is excessively sensitive to the outlier element. As we discuss in Section \ref{sec:synthetic1} and show in Tables \ref{tab:table_distance_hausdorff}--\ref{tab:table_distance_mj1}, only the MJ$_p$ family of semi-metrics properly identifies (\textbf{a}--\textbf{c}) as highly similar, with (\textbf{d}) the one set of structural breaks meaningfully different to the others.}
    \label{fig:Synthetic_time_series}
\end{figure}

First, we simulate a collection of time series $x_1,...,x_n$ from a GARCH model \citep{Lamoureux1990} with $m$ structural breaks determined by jumps at the points $\tau_1,...,\tau_m$. Each time series $x_i$  follows a Student-$t$ distribution with certain specified mean and variance functions. The mean function $\mu_t$ contains an autoregressive AR(1) process and a jump component; the latter is a product of a jump direction and magnitude with Bernoulli and gamma distributions, respectively. The variance function $\sigma^2_t$ contains several terms: an order one short-term component, a long-term persistence component, and a leverage effect component. We display four simulated simulated time series $x_i$ with specified sets of structural breaks $\tau_j$ in Figure \ref{fig:Synthetic_time_series}.

Next, we compute the distance matrix and associated affinity matrix between the four synthetic time series, relative to the Hausdorff metric, Wasserstein metric, and MJ$_{0.5}$ and MJ$_1$ semi-metrics, respectively, and display them in Tables \ref{tab:table_distance_hausdorff}--\ref{tab:table_distance_mj1}. These tables collectively illustrate the advantages of the MJ$_p$ semi-metrics compared with the Hausdorff and Wasserstein metrics first discussed in Section \ref{sec:theoreticalproperties}. First, the Wasserstein metric gives the lowest affinity score between the Time Series 1 and 2, which have 8 out of 9 of their structural breaks in common. These remarkably similar sets of structural breaks are given much lower distance and hence higher affinity under the MJ$_{0.5}$ and MJ$_1$ distances, illustrating Proposition \ref{prop:Wasserstein}. Next, the Hausdorff metric is far too sensitive to outliers; while TS1 and TS3 have 8 out of 9 points in common, these two time series are given the highest Hausdorff distance among the collection, hence an affinity equal to 0. Similarly, TS2 and TS3 have 7 of 9 points in common, but their assigned affinity is 0.1. The Wasserstein, MJ$_{0.5}$, and MJ$_1$ distances all recognize the similarity between TS1 and TS3 (as well as TS2 and TS3), with high affinity scores. Once again the MJ$_{0.5}$ and MJ$_1$ perform better than the Wasserstein in discerning the strong similarity between these time series. We provide the time series TS4 as a reference time series that is quite distinct in its structural breaks from TS1, TS2, and TS3. Only the  MJ$_{0.5}$ and MJ$_1$ assign TS1, TS2, and TS3 mutually high affinity scores, and only an algorithm using them to measure distances between sets of structural breaks would diversify away from including an unsuitably high quantity of TS1, TS2, and TS3 in one asset portfolio. Thus, this simple example of just four synthetic time series highlights the advantages of the MJ$_p$ semi-metric over the Hausdorff and Wasserstein metrics, illustrating the theoretical properties proven in Propositions \ref{prop:Hausdorff} and \ref{prop:Wasserstein}.

\vspace{-9pt}

\begin{table}[H]\caption{Hausdorff distance matrix and affinity matrices between four synthetic time series structural breaks. This allocates an unsuitably excessive distance between time series (\textbf{a},\textbf{c}) of Figure \ref{fig:Synthetic_time_series}, illustrating the Hausdorff metric's sensitivity to outliers.}
\label{tab:table_distance_hausdorff}\vspace{-3pt}
\begin{center}
\[
D=\begin{bmatrix}
0 & 100 & 1000 & 900 \\ 
100 & 0 & 900 & 800 \\
1000 & 900 & 0 & 900 \\
900 & 800 & 900 & 0 
\end{bmatrix}; A = 
\begin{bmatrix}
1 & 0.9 & 0 & 0.1 \\ 
0.9 & 1 & 0.1 & 0.2 \\
0 & 0.1 & 1 & 0.1 \\
0.1 & 0.2 & 0.1 & 1 
\end{bmatrix}
\]

\end{center}
\end{table}\vspace{-18pt}

\begin{table}[H]\caption{Wasserstein distance matrix and affinity matrices between four synthetic time series structural breaks. This allocates an unsuitably excessive distance between time series (\textbf{a},\textbf{b}) of Figure~\ref{fig:Synthetic_time_series}, despite their high intersection, showing an undesirable property when measuring discrepancy between sets.}
\label{tab:table_distance_wasserstein}\vspace{-3pt}
\begin{center}
\[
D=\begin{bmatrix}
0 & 100 & 111 & 933 \\ 
100 & 0 & 189 & 833 \\
111 & 189 & 0 & 844 \\
933 & 833 & 844 & 0
\end{bmatrix}; A = 
\begin{bmatrix}
1 & 0.89 & 0.88 & 0 \\ 
0.89 & 1 & 0.78 & 0.11 \\
0.88 & 0.78 & 1 & 0.10 \\
0 & 0.11 & 0.10 & 1 
\end{bmatrix}
\]

\end{center}
\end{table}\vspace{-18pt}

\begin{table}[H]\caption{$MJ_{0.5}$ distance matrix and affinity matrices between four synthetic time series structural breaks. This appropriately identifies (\textbf{a},\textbf{b}) as highly similar, (\textbf{c}) a little bit further away, and (\textbf{d}) the clear outlier element. }
\label{tab:table_distance_mj05}\vspace{-3pt}
\begin{center}
\[
D=\begin{bmatrix}
0 & 1 & 5 & 461 \\ 
1 & 0 & 13 & 306 \\
5 & 13 & 0 & 327 \\
461 & 306 & 327 & 0
\end{bmatrix}; A = 
\begin{bmatrix}
1 & 0.998 & 0.989 & 0 \\ 
0.998 & 1 & 0.972 & 0.336 \\
0.989 & 0.972 & 1 & 0.291 \\
0 & 0.336 & 0.291 & 1 
\end{bmatrix}
\]

\end{center}
\end{table}\vspace{-18pt}

\begin{table}[H]\caption{$MJ_{1}$ distance matrix and affinity matrices between four synthetic time series' structural breaks. This appropriately identifies (\textbf{a},\textbf{b}) as highly similar, (\textbf{c}) a little bit further away, and (\textbf{d}) the clear outlier element. }
\label{tab:table_distance_mj1}\vspace{-3pt}
\begin{center}
\[
D=\begin{bmatrix}
0 & 11 & 61 & 517 \\ 
11 & 0 & 72 & 417 \\
61 & 72 & 0 & 367 \\
517 & 417 & 367 & 0 
\end{bmatrix};  A = 
\begin{bmatrix}
1 & 0.98 & 0.88 & 0 \\ 
0.98 & 1 & 0.86 & 0.19 \\
0.88 & 0.86 & 1 & 0.29 \\
0 & 0.19 & 0.29 & 1 
\end{bmatrix}
\]

\end{center}
\end{table}\vspace{-24pt}

\subsection{Synthetic data: portfolio optimization experiments}
\label{sec:synthetic2}

\begin{figure}[H]
         \includegraphics[width=.8\textwidth]{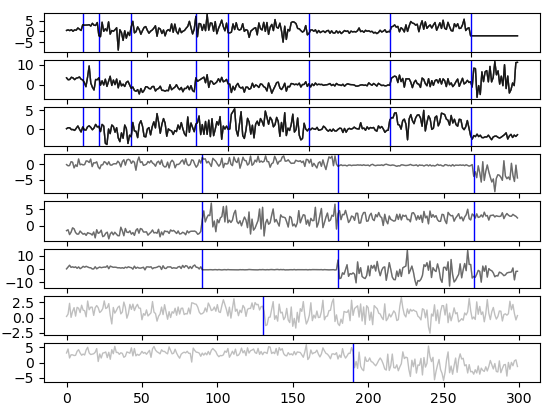}
     \caption{Synthetic time series with fixed historical returns and specified structural breaks, identifying distributional changes. As the historical returns are kept constant, only the sets of structural breaks are used in the objective function (\ref{eq:MJ}). These time series form the basis of the synthetic experiment in Section \ref{sec:synthetic2}. }
     \label{fig:Experiment_1_TimeSeries}
 \end{figure}

In this section, we apply our portfolio optimization methodology to synthetic data and illustrate the resulting allocation of assets. We generate eight synthetic time series with specified structural breaks. For simplicity, Assets 1--3 have identical numbers of change points with identical locations, as do Assets 4--6; Assets 7 and 8 are outliers. In addition, we set all time series to have the exact same historical return $\mathbb{E}R_i$, so that the numerator of (\ref{eq:MJ}) is a positive constant, regardless of the selection of weights. Thus, maximizing the MJ ratio (\ref{eq:MJ}) is equivalent to minimizing its denominator $\boldsymbol{w}^{T} A \boldsymbol{w}.$ We display the synthetic time series, together with structural breaks, in Figure \ref{fig:Experiment_1_TimeSeries}. We allocate our portfolio subject to a typical constraint  $5\% \leq w_i  \leq 40\%,   \forall i=1,...,8$. Such upper and lower bounds are frequently used in real-world investment policy statements \citep{RussellPolicy}, and may be tightened if desired.

Table \ref{tab:result_table_1} records the weights subject to the aforementioned constraints and conditions. The results demonstrate that our optimization framework is able to produce a more even distribution of change points across the portfolio. Assets 7 and 8, with significantly different breaks from the rest of the collection, are allocated more weight: 33.5\% and 30.7\%, respectively. Much less weight, 6.9\%, is allocated to Assets 1, 2, and 3, and just 5\% is allocated to Assets 4, 5, and 6. This experiment demonstrates that the algorithm provides diversification with regards to highly correlated structural breaks. Traditional mean variance portfolio optimization would be unable to do so.

\begin{table}[H]\caption{Portfolio allocation results for synthetic data experiment in Section \ref{sec:synthetic2}. Our method chooses high weights in Assets 7 and 8 to diversify away from the high similarity in structural breaks among Assets 1--6.}
\label{tab:result_table_1}

\setlength{\tabcolsep}{8.3mm}
\resizebox{\textwidth}{!}{\begin{tabular}{ p{2.5cm}p{4.5cm}p{2cm}}
  \toprule
  \textbf{Asset} & \textbf{Number of Change Points} & \textbf{Weights} \\
  \midrule
  Asset$_{1}$ & 8 &  6.9\% \\
  Asset$_{2}$ & 8 & 6.9\%  \\
  Asset$_{3}$ & 8 & 6.9\% \\
  Asset$_{4}$ & 3 &  5\% \\
  Asset$_{5}$ & 3 & 5\% \\
  Asset$_{6}$ & 3 & 5\% \\
  Asset$_{7}$ & 1 & 33.49\%\\
  Asset$_{8}$ & 1 & 30.7\%\\
  \bottomrule
\end{tabular}}
 
\end{table}

\section{Real data results}
\label{sec:realdata}

In this section, we apply our methodology to real financial data. We envisage this method being suitable in an asset allocation context, so we use indices and commodities as our underlying candidate investments. We are essentially simulating the role of an asset allocator, such as a pension fund or endowment, interested in macroeconomic asset allocation decisions. There are eight assets we allocate between to illustrate our method: the S\&P 500, Dow Jones Index, Nikkei 225 Index, BOVESPA Index, Stoxx 50 Index, ASX 200, oil spot price, and gold spot price, all between January 2009 and November 2019. There are several important details and assumptions in our experiments on real data:
\begin{enumerate}
    \item We train our algorithm over a relatively long period to estimate the true dynamics between various assets' structural breaks as precisely as possible. Training the algorithm on longer periods provides a more accurate assessment of similarity in varying market dynamics. 
    \item However, there is a balance between going back far enough to learn appropriate dynamics between asset classes and using too much history that relationships between assets no longer behave the way in which they were estimated. The behavior of individual asset classes and their relationships may change over time.
    \item The period from January 2018--June 2019 is a suitable out-of-sample period to test the algorithm, due to the varied market conditions. Most of 2018 provided relatively buoyant equity market returns, with a sharp drop in December 2018, followed by a prolonged recovery until June 2019. We wish to examine how candidate portfolios will perform in various market conditions, particularly in the presence of large drawdowns. In addition, we do not wish to test our algorithm during a period that is too similar to the training interval, as performance could be artificially strong. Thus, this is a suitable period to compare the optimization algorithms' performance.
    
    \item We did not include the COVID-19 market crisis in our test data to ensure that our training data have broadly similar dynamics to the out-of-sample data set. We include a targeted analysis of the COVID-19 crisis in Section \ref{sec:Covid}.

    \item The role of asset allocation is often guided by an investment policy statement that provides upper and lower bounds for capital allocation decisions. This is captured in the candidate weights' constraints. During pronounced bull and bear markets, institutional asset allocators may not have the flexibility to implement global optimization solutions. For example, if two asset classes had significantly higher returns and lower volatility than the remainder of candidate investments, the unconstrained solution would allocate all portfolio weight into these two assets. Investment weighting constraints prevent these contrived scenarios from occurring. For our constraints, we place a minimum 5\% and maximum 25\% of portfolio assets in any candidate investment. This is one of several typical constraints imposed in real-world policy statements---indeed, investment policy statements may include this as their only constraint \mbox{\citep{RussellPolicy}}.  As mentioned in Section \ref{sec:theoreticalproperties}, we may impose additional constraints by combining with other optimization methods cited in Section \ref{sec:intro}. 
    
    \item Our method provides an advantage over the simple correlation measure by addressing all three limitations in Section \ref{sec:framework}. One possible drawback to our proposed method, however, is that to learn meaningful relationships between assets' structural breaks, a long time series history is needed, preferably with many structural breaks observed. 
    
        \item When considering portfolio risk in an optimization framework, investors have a variety of measures they may choose to optimize over. Standard deviation, $\beta$, downside deviation, and tracking error are just several of these. Our CPO model introduces a mathematical framework that addresses peak-to-trough (drawdown) losses and erratic behavior as a measure of risk. Specifically, the model captures simultaneous asset shocks and aims to minimize the size of drawdowns by creating a uniform spread of change points across all portfolio holdings. We are unaware of any existing measure with these properties.

\end{enumerate}

\subsection{Training and validation procedure}
We train the algorithm between January 2009--December 2017 and test its performance on data from January 2018--June 2019. The training procedure learns the weights allocated to each candidate investment using the aforementioned objective function and constraints. We compare our change point optimization method (CPO) with nine other methods. These methods have been chosen as the best representation of comparative methods in the portfolio optimization literature. Given the breadth of research undertaken within the massive field of econometric 
portfolio optimization, no comparison with other methods can be completely exhaustive. This list is appropriate, as we judiciously chose methods that cover the most fundamental, best-known, and widely understood objective functions. They also include measures such as conditional value at risk, one of the fundamental measurements of tail risk in the field of econometrics. Our method has value in consciously pursuing an alternative mathematical attribute of diversification, which could also be incorporated with existing methods by portfolio averaging.

First, we apply the Mann--Whitney change point detection algorithm to the training data (log returns between January 2009--December 2017), identifying the locations of structural breaks in the mean for each possible asset. This yields eight sets of change points, where each point is indexed by time. Following Section \ref{sec:framework}, we apply the MJ$_{0.5}$ semi-metric to determine the distance between candidate sets of breaks. We optimize the MJ ratio objective function in (\ref{eq:MJ}) with respect to the weights, determining candidate weight allocations. Finally, we run an out-of-sample forecasting procedure using the weights estimated in our training data. We compare the predictive performance of the ten candidate methodologies between January 2018--June 2019. In addition, we apply agglomerative hierarchical clustering \citep{Mllner2013} to the resulting distance matrix between assets as exploratory analysis of their similarity with respect to structural breaks.

\begin{figure}[H]
        \includegraphics[width=.9\textwidth]{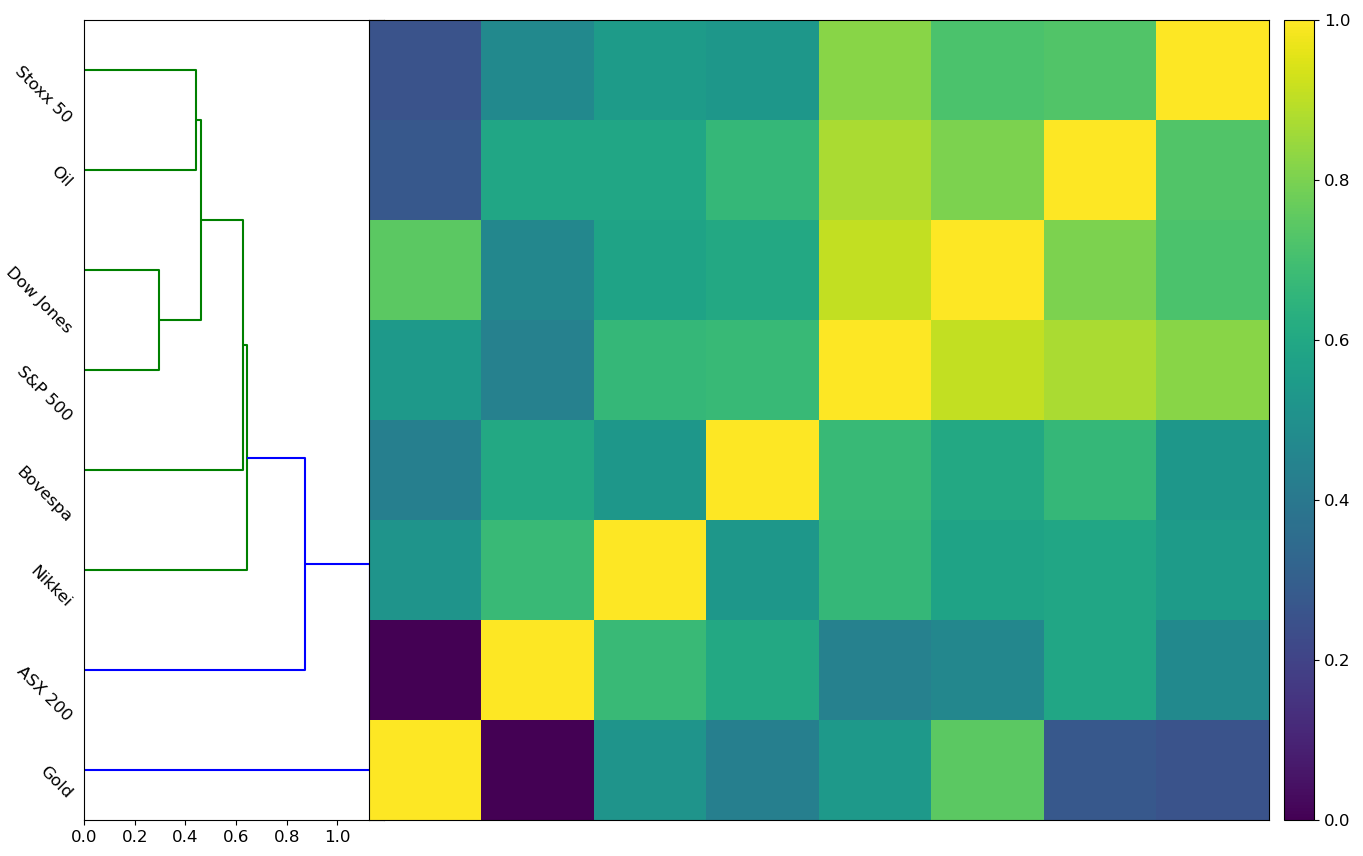}
    \caption{Hierarchical clustering applied to the distance matrix $D_{ij}$ between eight real assets' structural breaks, as defined in Section \ref{sec:framework}. Results indicate the structural breaks are most similar among the Dow Jones, S\&P 500, Stoxx 50, and oil. Gold is the most anomalous asset regarding structural break propagation.}
   \label{fig:Portfolio_analysis}
\end{figure}

This identifies a cluster of four highly similar assets (S\&P 500, Dow Jones, Stoxx 50, and oil), a cluster of three moderately similar assets (BOVESPA, Nikkei 225, and ASX 200), and an outlier in gold, displayed in Figure \ref{fig:Portfolio_analysis}. These results confirm financial intuition and documented relationships between asset classes, in particular gold's properties as a safe-haven asset. Both the S\&P 500 and Dow Jones Index are determined to be in the same cluster and are accordingly quite similar. Given that there is significant overlap in the constituents of both indices, this is a logical finding.

\subsection{Out-of-sample performance and distributional properties}
\label{sec:noncovidresults}

\begin{figure}[H]
    \begin{subfigure}[b]{0.65\textwidth}
        \includegraphics[width=\textwidth]{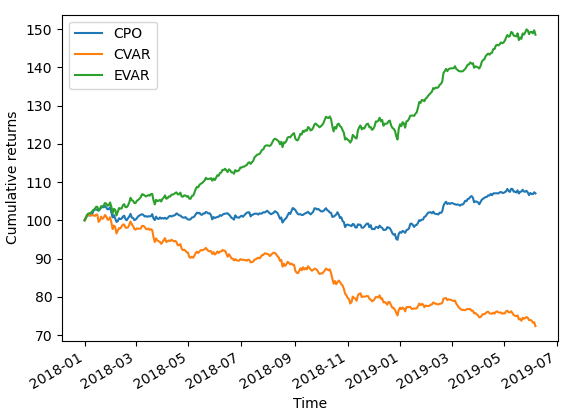}
        \caption{}
        \label{fig:Predictive_performance_3}
    \end{subfigure}\vspace{3pt}
    
    \begin{subfigure}[b]{0.65\textwidth}
        \includegraphics[width=\textwidth]{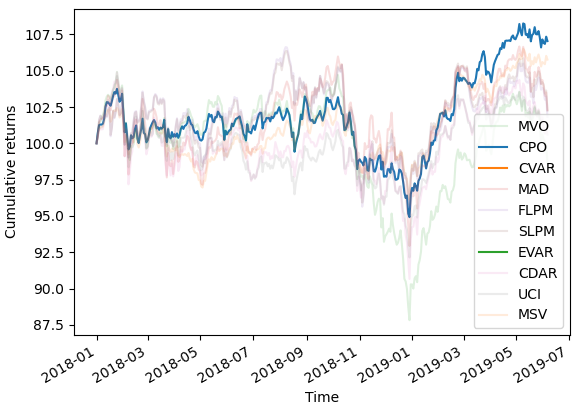}
        \caption{}
        \label{fig:Predictive_performance_bulk}
    \end{subfigure}\vspace{3pt}
    
    \begin{subfigure}[b]{0.65\textwidth}
        \includegraphics[width=\textwidth]{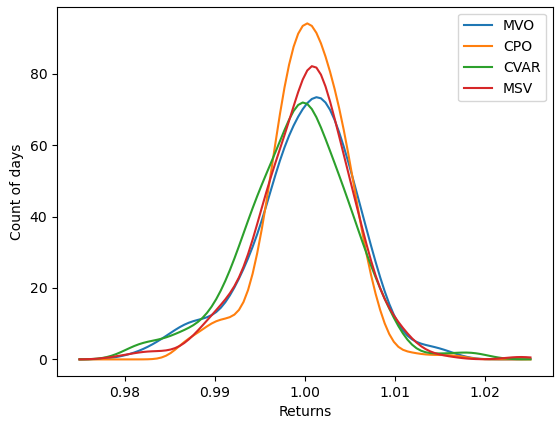}
        \caption{}
        \label{fig:Predictive_densities}
    \end{subfigure}
    \caption{Out-of-sample performance during the January 2018--June 2019 period comparing (\textbf{a}) returns of CPO with outliers CVAR and EVAR, (\textbf{b}) returns of all methods excluding CVAR and EVAR, and (\textbf{c})~distributions of MVO and CPO with commonly used methods MVO, MSV, and CVAR. Our methodology, CPO, generates the second-greatest returns, second-greatest Sharpe ratio, lowest volatility, and second-lowest drawdown of all methods.}
    \label{fig:Predictive_performance}
\end{figure}

Now, we compare all ten methods' out-of-sample performance, displaying cumulative returns over time in Figure \ref{fig:Predictive_performance} and documenting key metrics in Table \ref{tab:result_table_distribution}. With respect to cumulative returns, we see two clear outliers: the entropic value at risk (EVAR) and the conditional value at risk (CVAR) methods, which outperform and underperform, respectively.

\begin{table}[H]\caption{Results of our change point optimization (CPO) and nine other commonly used portfolio optimization methodologies applied to real data on the test period January 2018--June 2019. Utilized methods are mean-variance optimization (MVO), mean-semivariance (MSV), mean absolute deviation (MAD), first and second lower partial moment (FLPM and SLPM), conditional value at risk (CVAR), entropic value at risk (EVAR),  conditional drawdown at risk (CDAR), and Ulcer index (UCI). CPO and EVAR are the two best performing methods. CPO yields the lowest standard deviation, second-lowest drawdown and second-best return. Negative Sharpe ratios convey no useful information, so are omitted. For investors hoping to reduce portfolio volatility, the proposed CPO method is a suitable choice.}
\label{tab:result_table_distribution}

\setlength{\tabcolsep}{5mm}
\resizebox{\textwidth}{!}{\begin{tabular}{ m{1.1cm}m{2cm}m{2.0cm}m{1.3cm}p{1.8cm}m{1.4cm}}
 \toprule
 \textbf{Method}  & \textbf{Cumulative returns} & \textbf{Standard deviation} & \textbf{Sharpe ratio} & \textbf{Drawdown} & \textbf{Kurtosis} \\
\midrule
  CPO  & 107.04 &   \textbf{0.0045} & 0.99 & 8.83 & 1.06 \\ 
 MVO &   98.64 & 0.0060  & - & 17.08 & 1.54 \\
  MSV  & 105.76 &  0.0055 & 0.66 & \textbf{6.61} 
 & 1.61 \\ 
   MAD  & 102.28 &   0.0068 & 0.21 & 13.03 & 2.38 \\
 FLPM  & 101.82 &  0.0063 & 0.18 & 11.05 & \textbf{0.90} \\
 SLPM  & 102.26 &  0.0062 & 0.23 & 10.59 & 0.90 \\
  CVAR  & 72.32 & 0.0061 & - & 29.23 & 0.90 \\
 EVAR  & \textbf{148.55} & 0.0053 & \textbf{5.77} & 27.17 & 1.57 \\
 CDAR  & 100.66 &   0.0066 & 0.063  & 14.22 & 2.25 \\ 
 UCI  & 100.60 &  0.0055 & 0.069 & 12.47 & 1.61 \\
 \bottomrule
\end{tabular}}
 
\end{table}

For this purpose, Figure \ref{fig:Predictive_performance}a displays just EVAR, CVAR, and our proposed change point optimization method (CPO). We see that CPO produces more stable return trajectories than each outlier method, albeit less total returns than EVAR. Subsequently, we exclude EVAR and CVAR and compare the remaining eight methods in Figure \ref{fig:Predictive_performance}b. Among these methods, the CPO method generates the greatest cumulative returns and the lowest standard deviation. For a candidate investor most focused on generating significant returns with minimal volatility, CPO exhibits the most favorable risk--return profile. 

In Figure \ref{fig:Predictive_performance}c, we contrast the density of daily returns for CPO and three commonly applied portfolio optimization methodologies: mean--variance optimization (MVO), mean--semivariance (MSV), and CVAR. The thinner tails exhibited by CPO show that this method provides consistently reduced volatility in returns. Together with the fact that CPO provides the highest cumulative returns of these methods, we see CPO provides a superior risk-adjusted return compared with these comparable competitive measures. This makes it the most desirable portfolio among all non-outlier candidate methods.

All methods' cumulative returns are documented precisely in Table \ref{tab:result_table_distribution}. CPO produces the second-highest cumulative returns (and second-highest Sharpe ratio) when examined on our test data, with a final value of 107.04. The best-performing method is EVAR, which generates a final value of 148.55. Interestingly, the EVAR method also exhibits the highest volatility (standard deviation) and one of the highest drawdowns. This suggests that although CPO did not produce the highest returns, it may have still been the most preferred methodology for investors concerned with volatility.

Next, we examine the out-of-sample standard deviation performance among all comparative methods. CPO produces the lowest standard deviation, with a score of 0.0045. The second-best-performing method is the MSV, with a standard deviation of 0.0055, and the worst-performing methods are the mean absolute deviation method (MAD) and EVAR. That is, despite the outlier method's performance in returns, it exhibits significant volatility.

CPO also performs very strongly with respect to portfolio drawdown. CPO produces the second-lowest drawdown, with a total score of 8.83. The best-performing comparative method is MSV, with a drawdown of 6.61, while CVAR and EVAR have the most significant drawdowns of 29.23 and 27.17, respectively. Again, for investors concerned with strong performance and minimal risk---the CPO and MSV methods produce the most favorable profiles. High levels of drawdown can be particularly concerning for portfolio managers, as respective clients actively tracking their investments may panic during drawdowns and request a withdrawal of funds. For active managers, this is particularly concerning, as there is a heightened chance of funds dropping below the high watermark. 

Finally, we turn to kurtosis, where CPO exhibits a level that is approximately average among all the examined methodologies. Compared with other methods that generate strong returns profiles such as EVAR and MSV,  CPO has a kurtosis of 1.06, while EVAR and MSV produce scores of 1.57 and 1.61, respectively. Given the positive skew in the CPO distribution, the suppressed kurtosis value is likely indicative of less tail risk in the CPO predictive distribution. The best-performing methods are CVAR and the first and second lower partial moment methods (Omega and Sortino ratios, respectively), all producing scores of 0.9. Although this is indicative of less tail risk, the kurtosis scores of the predictive distribution are likely lower due to reduced average daily returns.

\subsection{Performance during COVID-19}
\label{sec:Covid}

\begin{figure}[H]
    \includegraphics[width=.9\textwidth]{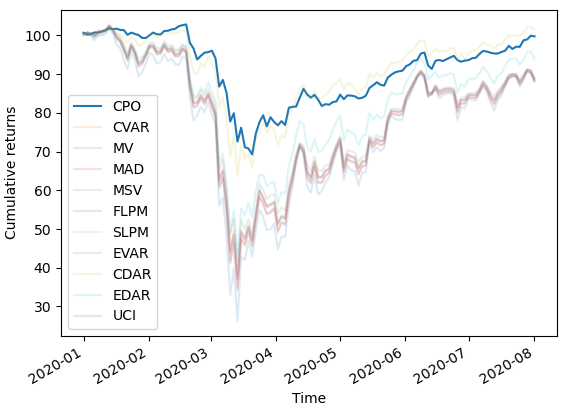}
    \caption{Out-of-sample performance during the height of the COVID-19 market crisis: January--August 2020. We compare the returns of CPO with all methods. Our methodology, CPO, generates the second-greatest returns, lowest volatility, lowest kurtosis, and lowest drawdown of all methods.}
    \label{fig:COVID_experiment}
\end{figure}

We devote this section specifically to an analysis of the performance of our methodology (and several others) during the market crash associated with the onset of the COVID-19 pandemic in early 2020 \citep{Akhtaruzzaman2020,Okorie2020,Jamesfincovid}. We focus on the period of January--June 2020, which was associated with a significant market decline and then correction, including some of the worst one-day market drops in history \citep{CNBC}. We use the same methods as in the previous Section \ref{sec:noncovidresults} and investigate the performance of the corresponding portfolios, renormalized to begin at 100\% as of the beginning of January 2020. We display cumulative returns over time in Figure \ref{fig:COVID_experiment} and document the same metrics as before in Table \ref{tab:result_table_distribution_COVID}.

Similar to Section \ref{sec:noncovidresults}, we see again that CPO produces more stable return trajectories than any other method (albeit with CDAR ultimately producing higher returns at the end of the period). CPO produces the second-highest cumulative returns when examined on our COVID-19 data, with a final value of 99.76. Only CDAR produces ultimately higher returns (101.43), at the cost of higher volatility, drawdown, and kurtosis. Again, for a candidate investor focused on weathering a market crisis, CPO exhibits the most favorable risk--return profile. We remark that all optimization methods under investigation share a broadly similar cumulative returns trajectory. This suggests that, regardless of the weight allocations (which vary meaningfully across methods), completely avoiding the market crash is virtually impossible in a long-only setting. One can see that our proposed methodology does an excellent job at mitigating downside risk during a crisis, while maintaining solid positive returns during the subsequent market recovery.

In particular, CPO produces the best (minimal) volatility, drawdown, and kurtosis among all methods, as seen in Table \ref{tab:result_table_distribution_COVID}. Indeed, CPO carries the lowest standard deviation of 0.018, while the second-lowest standard deviation is 0.025. As in Section \ref{sec:noncovidresults}, the suppressed kurtosis of CPO is likely indicative of less tail risk in returns distribution. Combined with the minimal drawdown, this is an especially welcome feature for investors seeking a portfolio to best weather a market crisis, where excessive drawdowns can cause panic and bank runs.

\begin{table}[H]
 \caption{Results of our change point optimization (CPO) and nine other commonly used portfolio optimization methodologies applied to real data during the height of the COVID-19 market crisis: January--August 2020. Utilized methods are mean--variance optimization (MVO), mean--semivariance (MSV), mean absolute deviation (MAD), first and second lower partial moment (FLPM and SLPM), conditional value at risk (CVAR), entropic value at risk (EVAR),  conditional drawdown at risk (CDAR), and Ulcer index (UCI). CPO and CDAR are the two best-performing methods. CPO yields the lowest volatility, drawdown, and kurtosis, and the second best return. For investors hoping to reduce portfolio volatility, the proposed CPO method is a suitable choice. As almost all returns are negative, the Sharpe ratio is uninformative and has been omitted.}
\label{tab:result_table_distribution_COVID}

\setlength{\tabcolsep}{3.3mm}
\resizebox{\textwidth}{!}{\begin{tabular}{ p{1.3cm}p{3.4cm}p{3.4cm}p{1.8cm}p{1.4cm}}
 \toprule
 \textbf{Method}  & \textbf{Cumulative returns} & \textbf{Standard deviation}   & \textbf{Drawdown} &\textbf{Kurtosis} \\
 \midrule
  CPO  & 99.76 &   \textbf{0.018}  & \textbf{33.55} & \textbf{8.55} \\
 MVO &   88.82 & 0.036   & 65.13 & 15.48 \\
  MSV  & 88.69 &  0.036  & 65.54 & 15.60 \\
   MAD  & 88.06 &   0.038  & 67.71 & 15.99 \\
 FLPM  & 88.08 &  0.038  & 68.39 & 16.11 \\
 SLPM  & 88.69 &  0.036  & 65.51 & 15.60 \\
 CVAR  & 88.45 & 0.035  & 64.41 & 15.54 \\
 EVAR  & 89.12 & 0.033  & 60.99 & 14.90 \\
 CDAR  & \textbf{101.43} &   0.025   & 38.38 & 9.49 \\
 UCI  & 88.40 &  0.045  & 75.58 & 17.15 \\
 \bottomrule
\end{tabular}}
 
\end{table}

\subsection{Sampling study of structural breaks between countries' financial indices}

In this section, we conduct a sampling study to investigate patterns in the collective distance between financial assets. For robustness, we draw data from a completely independent selection of assets as the prior sections, analyzing log returns data of the national financial indices of 19 countries. Our chosen countries are Australia, Brazil, Canada, China, France, Germany, India, Indonesia, Italy, Japan, Korea, the Netherlands, Russia, Saudi Arabia, Spain, Switzerland, Turkey, the United Kingdom (UK), and the United States (US), with data ranging from 2001--2020. We begin by applying our change point algorithm (described in Appendix \ref{Appendix_CPD}) to obtain sets of structural breaks $S_i, i=1,...,19$ for each country.

\begin{figure}[H]
    \includegraphics[width=.95\textwidth]{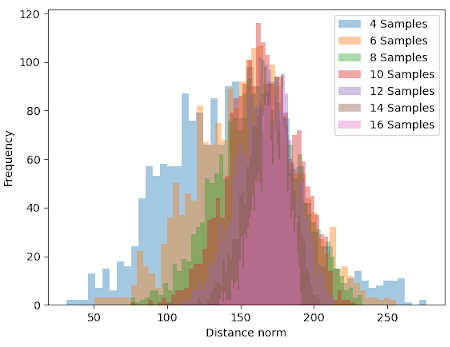}
    \caption{Distributions for the normalized distance matrix norm $\|D\|$ when repeatedly drawing $K=2000$ samples of size $n$ from our collection of 19 countries and computing the distances between structural breaks. The range of values increases but tightens in variance as $n$ increases, reflecting that the normal benefits of diversification when increasing portfolio size also apply to the quantitative measure of distances between structural breaks.  }
    \label{fig:sampling}
\end{figure}

\textls[-25]{Next, we perform a variety of repeated sample experiments. We vary $n=4,6,8,10,12,14,16$} and for each value of $n$, sample $K=2000$ draws of $n$ countries (without replacement) from the collection. A draw of $n$ countries produces an $n \times n$ distance matrix $D$ (using the MJ$_{0.5}$ semi-metric as elsewhere in the paper) between the countries. We compute the following normalized $L^1$ norm of the matrix $D$ to measure the collective magnitude of all distances between countries:
\begin{align}
    \label{eq:norm}
    \|D\|=\frac{1}{n(n-1)} \sum_{i,j=1}^n |D_{ij}|.
\end{align}
We normalize by the number of nonzero elements in this distance matrix, $n(n-1)$. Thus, the sampling procedure produces $K=2000$ values of $\|D\|$ for each size $n$. In Figure \ref{fig:sampling}, we show the distributions of values for each $n$. We also record a 90\% interval consisting of the 5th and 95th quantile in Table \ref{tab:sampling_confidenceintervals}. As $n$ increases, we observe an increase in the overall mean of the distribution---this effect is relatively rapid at first but slows after $n=8$. We also observe a relatively quick increase in the lower limit and a slower decrease in the upper limit. This is to be somewhat expected. In small samples (such as $n=4$), it is possible to repeatedly by chance select just four countries that are relatively similar to each other in terms of structural breaks. However, selecting 8 or 10 countries that are all similar to each other (and hence yielding a small value of $\|D\|$) is much less likely, producing a greater value of the lower limit.

\vspace{-9pt}

\begin{table}[H]

\caption{Ninety percent confidence intervals for the normalized distance matrix norm $\|D\|$ when repeatedly drawing $K=2000$ samples of size $n$ from our collection of 19 countries and computing the distances between structural breaks. The range of values increases but tightens in variance as $n$ increases, reflecting that the normal benefits of diversification when increasing portfolio size also apply to the quantitative measure of distances between structural breaks.}
\label{tab:sampling_confidenceintervals}

\setlength{\tabcolsep}{9.5mm}
\resizebox{\textwidth}{!}{\begin{tabular}{ p{2.5cm}p{3cm}p{3cm}}
 \toprule
 \textbf{Sample size}  & \textbf{Lower limit} & \textbf{Upper limit}  \\
 \midrule
  4  & 72.99 &   213.17  \\
 6 &   100.70 & 210.92  \\
  8 & 113.35 &  208.32 \\
   10  & 127.02 &  201.08  \\
 12  & 135.86 &  196.36  \\
 14  & 143.81 &  191.59  \\
 16  & 151.42 & 186.31 \\
  \bottomrule
\end{tabular}}
 
\end{table}

The next key finding in this experiment is the significant reduction in the spread of the distribution of matrix norms, as a larger number of countries are sampled. The light blue coloring, which is a relatively diffuse distribution, corresponds to when only four country indices are sequentially sampled. This distribution is centered $\sim$150, with a total range of approximately 250. As the number of stocks sampled increases, we see the distribution of norm values become successively narrower (exhibiting less variance). This culminates in the pink distribution ($n=16$), centered around $\sim$175, with a total range of approximately~50. 

\section{Discussion}
\label{sec:conclusion}
We  proposed a novel optimization method, which utilizes semi-metrics between sets of structural breaks, to reduce simultaneous asset shocks across an investment portfolio. This should be understood as an alternative approach to avoid simultaneous losses that can gravely threaten an investor's holdings. Experiments on synthetic data confirm that we are able to detect similar time series in terms of structural breaks and accordingly allocate highly similar assets less portfolio weight. In addition, synthetic experiments illustrate concretely the proven theoretical properties of our particular choice of semi-metric and its benefits over other existing metrics between finite sets. Experiments on real data suggest that our method may significantly reduce both portfolio volatility and drawdown when compared with numerous existing methodologies. 

This novel optimization framework may have significant implications for asset allocation and portfolio management professionals who are interested in alternative measures of risk. Our method diversifies well away from portfolio drawdown and seeks to avoid the erratic behavior of highly clustered change points. Our method is flexible, and different change point algorithms may be married with other distance measures or objective functions for alternative approaches. Finally, our method's efficient implementation and reasonable complexity mean it easily generalizes to large instances of the portfolio optimization problem, especially when convex real-world constraints are applied. In the event of more difficult constraints, the calculation of our distances between sets of change points could be combined with efficient optimization approaches in the econometrics literature.

There are several limitations in our optimization framework. Change point detection methodologies vary widely, and there is substantial literature on their advantages and disadvantages \citep{Hawkins2005,gustafsson2001}. One potential limitation of our chosen change point detection algorithm is its deterministic selection (via hypothesis testing and maximization of test parameters) of change point locations. Alternative approaches, such as Bayesian methods, may provide a probabilistic approach incorporating the uncertainty around change points' existence. Next, our methodology requires a long training period to learn meaningful relationships between assets' structural breaks, and it is conceivable that such relationships from asset histories no longer hold in the present. Furthermore, any distance measure between finite sets will have its limitations, such as the Hausdorff metric's sensitivity to outliers, the semi-metrics' failure in the triangle inequality, or the potentially excessive averaging in the MJ$_p$ family. To ameliorate these limitations, we believe model averaging with other methods that require a smaller training time and other mathematical quantities other than distances between finite sets could be beneficial. Furthermore, it is our hypothesis that structural breaks are more likely to have an underlying persistence that is more consistent than their returns. Each asset's structural break propagation most likely has a strong link to the asset's existence within the very complex system of the global economy. Structural change in this manner is likely to occur at a much lower frequency than drivers which dictate changes in market returns and volatility (such as market sentiment).

Future research could work to ameliorate such limitations by building on the proposed framework. Different change point detection algorithms could be used for other stochastic properties, such as the variance in the returns, or to reflect uncertainty in the breaks' existence. One could explore how results change when the order of $p$ changes within the MJ$_p$ semi-metric, or when entirely different distance measures are used between sets. One could conceivably calibrate the value of $p$, such as selecting a portfolio based on the optimal MJ ratio for various ratios of $p$, and then using another optimization function such as the Sharpe ratio to tune $p$ as a hyperparameter. More broadly, it is conceivable that one could diversify between sets through other means than reducing their discrepancy to a single scalar value. Even with the judicious choice of $p$, the outlier sensitivity in the denominator $\max D$ of (\ref{eq:affinity}) is a limitation of our framework, acknowledged in Remark \ref{remark:outliers}, so alternative constructions other than this affinity matrix could be explored. Furthermore, one could combine our methodology, which requires a long history of training data, with alternative methods that give more weight to the recent behavior of a financial asset.

This work not only has broad theoretical value, offering for the first time how structural breaks may be used in portfolio optimization and proving numerous properties of our implementation relative to other discrepancy measures between structural breaks, but it also has experimental value beyond our specific application. Model stacking and averaging, also known as ensemble learning \citep{Sagi2018_ensemble}, has proven to have great utility in portfolio optimization \citep{nl2021_ensemble,Shen2019_ensemble}, even combining rather unrelated methods. As we are unaware of portfolio techniques that study structural breaks, we believe model stacking with other methods (focused on variance, tail risk, conditional correlations, and others) could benefit by bringing in another mathematical property of financial time series to assist in diversification and econometric-derived reduction in risk. Alternatively, our method could be combined with numerous methods to clean noise in financial time series prior to analysis. Indeed, for the standard Markowitz model of covariance and correlations, numerous authors have argued that correlation matrices approximate random matrices and should be cleaned \citep{Laloux1999,Wtorek2020}, with numerous approaches proposed for this purpose \citep{Bun2017}. Conceivably, such preprocessing could be applied in our setting to the sets of structural breaks of different~assets.

\section{Conclusion}
\label{sec:newconclusion}
We  proposed a new concept of using distances between structural breaks of time series for portfolio optimization and provided a specific implementation. Our first implementation has promising results in its own right and also offers numerous directions for further research by incorporating other advances in the literature, such as noise cleaning and model stacking. More broadly, we hope this paper will invite other novel approaches and concepts towards exploring different ``root causes'' of simultaneous shocks in financial holdings. Avoiding these has value not just for investors and portfolio managers but for the health of the broader economy as a whole.

\vspace{6pt}

\authorcontributions{N.J. and M.M. are equal first authors, playing an equal role in every aspect of the manuscript. J.C. performed simulation experiments and provided edits to the manuscript. All authors have read and agreed to the published version of the manuscript.}

\funding{This research received no external funding.}

\institutionalreview{Not applicable.}

\informedconsent{Not applicable.}

\dataavailability{Data analyzed in this study were obtained from Bloomberg.}

\acknowledgments{M.M. would like to thank Xiao Ting of Tsinghua Sanya International Mathematics Forum for making the time in China easy and pleasant.}

\conflictsofinterest{The authors declare no conflict of interest.}

\abbreviations{Abbreviations}{
The following abbreviations are used in this manuscript:\\

\noindent 
\begin{tabular}{@{}ll}
CPO & Change point optimization method \\
MVO & Mean--variance optimization\\
MSV & Mean--semivariance \\
MAD & Mean absolute deviation\\
FLPM & First lower partial moment\\
SLPM & Second lower partial moment \\
CVAR & Conditional value at risk \\
EVAR & Entropic value at risk\\
CDAR & Conditional drawdown at risk \\
UCI & Ulcer index \\
CPM & Change point model
\end{tabular}
}

\appendixtitles{yes} 
\appendixstart
\appendix
\section[\appendixname~\thesection]{Change point detection algorithm}
\label{Appendix_CPD}

In this section, we describe the change point detection algorithm used in Section \ref{sec:framework}. The general change point detection framework is the following: a sequence of observations $x_1,x_2,...,x_n$ are drawn from random variables $X_1, X_2,...,X_n$ and undergo an unknown number of changes in distribution at points $\tau_1,...,\tau_m$. We assume observations are independent and identically distributed between change points, that is, between each change point, a random sampling of the distribution is occurring. \cite{RossCPM} notates this as follows:\begin{align}
    X_{i} \sim 
    \begin{cases}
      F_{0} \text{ if } i \leq \tau_1 \\
      F_{1} \text{ if } \tau_1 < i \leq  \tau_2  \\
      F_{2} \text{ if } \tau_2 < i  \leq \tau_3,  \\
      \hdots
    \end{cases}
\end{align}
While this requirement of independence may appear restrictive, dependence can generally be accounted by several means, such as modeling the underlying dynamics or drift process and then applying a change point algorithm to the model residuals or one-step-ahead prediction errors, as described by \cite{gustafsson2001}. The change point models described below that we apply in this paper follow \cite{RossCPM}.

\subsection{Batch detection (Phase I)}
This first phase of change point detection is retrospective. We are given a finite sequence of observations $x_1,\ldots,x_n$ from random variables $X_1,\ldots,X_n$. For simplicity, we assume at most one change point exists. If a change point exists at time $k$, this means observations have a distribution of $F_0$ prior to the change point and a distribution of $F_1$ proceeding the change point, where $F_0 \neq F_1$. Then, one must test between the following two hypotheses for each $k$: 

\begin{align}
    H_0: X_{i} \sim F_0, i = 1,...,n
\end{align}

\begin{align}
    H_1: X_{i} \sim 
    \begin{cases}
      F_{0} & i = 1,2,...,k \\
      F_{1}, & i = k + 1, k+2, ..., n  \\
    \end{cases}
\end{align}
and select the most suitable $k$.

One proceeds with a two-sample hypothesis test, where the choice of test depends on the assumptions about the underlying distributions. Nonparametric tests can be chosen to avoid distributional assumptions. One appropriately chooses a two-sample test statistic $D_{k,n}$ and a threshold $h_{k,n}$. If $D_{k,n}>h_{k,n}$, then the null hypothesis is rejected and one provisionally assumes that a change point has occurred after $x_k$. These test statistics $D_{k,n}$ are normalized to have mean $0$ and variance $1$ and are evaluated at all values $1 < k < n$; the largest value is assumed to be coincident with the existence of our sole change point. The test statistic is then
\begin{align}
    D_{n} = \max_{k=2,...,n-1} D_{k,n} = \max_{k=2,...,n-1} \Bigg| \frac{\Tilde{D}_{k,n} - \mu_{\Tilde{D}_{k,n}}}{\sigma_{\Tilde{D}_{k,n}}}  \Bigg|
\end{align}
where $\Tilde{D}_{k,n}$ are non-normalized statistics.

The null hypothesis of no change is rejected if $D_{n} > h_n$ for an appropriately chosen threshold $h_n$. In this case, we conclude that a (unique) change point has occurred, and its location is the value of $k$ which maximizes $D_{k,n}$. That is,
\begin{align}
    \hat{\tau} = \argmax_k D_{k,n}.
\end{align}
This threshold $h_n$ is chosen to bound the Type 1 error rate, as is commonplace in statistical hypothesis testing. First, one specifies an acceptable level $\alpha$ for the proportion of false positives, that is, the probability of falsely declaring that a change has occurred when in fact it has not. Then, $h_n$ is chosen as the upper $\alpha$ quantile of the distribution of $D_n$ under the null hypothesis. For the details of computation of this distribution, one can see \cite{RossCPM}. 

The computational cost of this first phase is $O(n)$, where $n$ is the number of observations. Indeed, this calculates and compares $n-2$ values of the (normalized) test statistics $D_{k,n}, k=2,...,n-1$. This is implemented efficiently in C\texttt{++}.

\subsection{Sequential detection (Phase II)}
In this second phase, the sequence $(x_t)_{t \geq 1}$ does not have a fixed length. New observations are continually received over time, and multiple change points may be present. Assuming no change point exists so far, this approach treats $x_1,..., x_t$ as a fixed-length sequence and computes $D_t$ as described in phase I. A change is flagged if $D_t > h_t$ for an appropriately chosen threshold. If no change is detected, the next observation $x_{t+1}$ is brought into the sequence of consideration. If a change is detected, the process restarts from the data point immediately following the detected change point. Thus, the procedure consists of a repeated sequence of hypothesis tests.

In this sequential setting, $h_t$ is selected so that the probability of incurring a Type 1 error is constant over time, so that under the null hypothesis of no change, the following~holds:
\begin{align}
    P(D_1 > h_1) &= \alpha,\\
    P(D_t > h_t | D_{t-1} \leq h_{t-1}, ... , D_{1} \leq h_{1}) &= \alpha, \ t > 1.
\end{align}
In this case, assuming that no change occurs, the expected number of observations received before a false positive detection occurs is equal to $\frac{1}{\alpha}$. This quantity is often referred to as the average run length, or ARL$_0$. Additional details on appropriate values of $h_t$ are detailed by \cite{RossCPM}.

In the context of our paper, this algorithm is performed on time series of length $T$. Thus, this second phase involves up to $T$ implementations of Phase I, with a new observation $x_{t+1}$ brought in each step. As the complexity cost of Phase I is up to $O(T)$, this means the total cost of the CPM algorithm is $O(T^2)$. This is implemented efficiently in C\texttt{++}.

\section[\appendixname~\thesection]{Overview and properties of distances between sets}
\label{Appendix_distances}

In this section, we provide an overview of (semi-)metric distances, with a focus on distance between (finite) sets, and motivate the choice of distance between sets of structural breaks chosen in (\ref{eq:MJdefn}) and (\ref{eq:MJchoice}).

\subsection{Overview of metrics}

We first recall the definition of a metric $d$ on a set $X$. A pairing $d: X \times X \to \mathbb{R}$ is called a metric if it satisfies the following axioms for all $x,y,z \in X$:
\begin{enumerate}
    \item $d(x,y) \geq 0$, with equality if and only if $x=y$;
    \item $d(x,y) = d(y,x)$;
    \item $d(x,z) \leq d(x,y)+d(y,z)$.
\end{enumerate}
$d$ is a \emph{semi-metric} if it satisfies (i) and (ii) but not necessarily (iii), which is known as the \emph{triangle inequality}. If $d$ is a metric on $X$, then the pair $(X,d)$ is called a metric space \mbox{\citep{RudinPMI}.}

As discussed in Section \ref{sec:motivation}, the focus of our methodology is measuring discrepancy between finite sets. With this in mind, the relevant class of (semi-)metrics for this paper is that between subsets of a given metric space. We begin with more explanation of a concept used in Section \ref{sec:framework}. Let $S$ be a subset of a metric space $X$, and $x \in X$. Then, the distance from the element $x$ to the set $S$ is defined as the minimal distance from $x$ to any point in $S$, computed as follows:
\begin{equation}
d(x,S) = \inf_{s \in S} d(x,s). \label{eq:min distance defn}
\end{equation}
Now $d(x,S) \geq 0$ with equality if and only if $x$ lies in the closure of $S$. In addition, $d(-,S); X \to \mathbb{R}$ is continuous. This quantity $d(x,S)$ is the base ingredient of several existing and recently introduced (semi-)metrics between sets.

\subsection{Distances between sets}

Now, let $S, T \subset X$ be (finite) subsets of any metric space. A common first notion of distance between these subsets is defined as the minimal distance between these subsets, defined by
\begin{equation}
    d_{\text{min}}(S,T) = \inf_{s \in S} d(s,T) =  \inf_{s \in S} \inf_{t \in T} d(s,t) = \inf_{s \in S, t \in T} d(s,t). \label{eq:min min distance}
\end{equation}    
Note $d_{\text{min}}(S,T) = 0$ if $S,T$ intersect. In fact, $d_{\text{min}}(S,T)=0$ if and only if their closures (in the ambient space $X$) intersect. So, this is not an effective metric between subsets, as it can frequently be zero for sets that are markedly different. We proceed to outline some existing (semi-)metrics between finite sets that have been used for various applications \citep{Conci2017}. 

We begin with the Hausdorff distance, already defined in Definition \ref{def:Hausdorff}:
\begin{align}
    d_{H}(S,T) =&  \max \left( \sup_{s \in S} d(s,T), \sup_{t \in T} d(t,S) \right) \\
    =&  \sup \{ d(s,T), s \in S; d(t,S), t \in T \}.
\end{align}
Essentially, the Hausdorff metric considers how separated $S$ and $T$ are at the most, rather than at least, compared with (\ref{eq:min min distance}). More precisely, it is the supremum or $L^{\infty}$ norm of all minimal distances from points $s \in S$ to $T$ and points $t\in T$  to $S$, as defined in (\ref{eq:min distance defn}). The Hausdorff distance satisfies the triangle inequality (so it is a true metric, rather than just a semi-metric), but this supremum is highly sensitive to even a single outlier. Indeed, this is the content of Proposition \ref{prop:Hausdorff} and Corollary \ref{cor:Hausdorff}, that just one element of $S$ or $T$ can result in great changes to $d_H(S,T)$.

Next, we discuss the pre-existing modified Hausdorff distances, which are semi-metrics that were used in computer vision and other tasks.

\begin{Definition}[Modified Hausdorff distance 1]
The first modified Hausdorff distance MH$_1$ is defined as follows \citep{Dubuisson1994,Deza2013}:
\begin{align}
\label{eq:MH1}
d^{\text{MH}}_1(S,T) = \max \left( \frac{1}{|S|} \sum_{s \in S} d(s,T), \frac{1}{|T|} \sum_{t \in T} d(t,S) \right).
\end{align}
\end{Definition}
\noindent It takes a first step at replacing the max in the Hausdorff distance with geometric averaging.

\begin{Definition}[Modified Hausdorff distance 2]
The second modified Hausdorff distance MH$_2$ is defined as follows \citep{Dubuisson1994,Eiter1997}:
\begin{align}
d^{\text{MH}}_2(S,T) = \sum_{s \in S} d(s,T) + \sum_{t \in T} d(t,S).
\end{align}
\end{Definition}
\noindent Unlike (\ref{eq:MH1}), \textls[-15]{this captures the total deviation between one set and another, with no averaging.}

\begin{Definition}[Modified Hausdorff distance 3]
The third modified Hausdorff distance MH$_3$ is defined as follows \citep{Dubuisson1994,Deza2013}:
\begin{align}
d^{\text{MH}}_3(S,T) = \frac{1}{|S|+|T|} \left(\sum_{s \in S} d(s,T) + \sum_{t \in T} d(t,S) \right).
\end{align}
\end{Definition}
\noindent This is a variant of (\ref{eq:MH1}) with a different averaging component, referred to as geometric mean error between two images. 

In addition to the Hausdorff metric and the three pre-existing modified Hausdorff distances defined above, there is also the Wasserstein distance, discussed in Section \ref{sec:theoreticalproperties}.

Now, we turn to the more recently introduced family of semi-metrics introduced in \cite{James2020_nsm} and motivate it as our choice for this paper. Here, we first introduced the MJ$_1$ semi-metric: 
\begin{align}
\label{eq:MJ1}
    d^1_{MJ}({S},{T}) = \frac{1}{2} \left(\frac{\sum_{t\in T} d(t,S)}{|T|} + \frac{\sum_{{s} \in {S}} d(s,T)}{|S|} \right).
\end{align}
Our initial motivation for this is as follow: In \cite{Dubuisson1994}, the authors asserted that their distance MH$_1$ is the best for image matching. To reach this conclusion, they took two steps. First, they compared three favorable operators, $f_2,f_3,$ and $ f_4$, each operating on minimal distances $d(s,T),d(t,S)$, as defined in Equation (\ref{eq:min distance defn}). They briefly argued that $f_2$, equivalent to taking the max in the MH$_1$, is preferable to other operators, citing a ``larger spread'.' Second, they argued that a process of averaging distances is superior to taking $K$th ranked distances, such as the median. We differed with the first step of their reasoning and replaced the max in their MH$_1$ with the $L^1$ norm average of all the minimum distances from $S$ to $T$ and $T$ to $S$, as seen in (\ref{eq:MJ1}). We proceeded to detail some reasons why we preferred the MJ$_1$ over the three aforementioned modified Hausdorff distances, explained in Propositions 3.1, 3.2, 3.5, and 3.6 of \cite{James2020_nsm}.

Regarding the second step of their reasoning \citep{Dubuisson1994}, we agreed that an averaging process was more suitable regarding outlier error than the alternative processes and chose to generalize this by using other $L^p$ norm averages. Thus, we introduced the family of MJ$_p$ semi-metrics. We define the MJ$_p$ distance by
\begin{align}
\label{eq:New MJ distance}
    d^p_{MJ}({S},{T}) = \Bigg(\frac{\sum_{t\in T} d(t,S)^p}{2|T|} + \frac{\sum_{{s} \in {S}} d(s,T)^p}{2|S|} \Bigg)^{\frac{1}{p}}.
\end{align}
The normalization within the expression is chosen such that 
\begin{align}
    d^p_{MJ}({S},{T}) \leq d_H(S,T) \text{ for all } p, \text{ and } \lim_{p \to \infty} d^p_{MJ}({S},{T}) = d_H(S,T).
\end{align}
Thus, $d_H$ can now be viewed as the $L^\infty$ norm of these distances, that is, our family of semi-metrics includes the Hausdorff distance as a limiting case when $p \rightarrow \infty$. So, the existing Hausdorff metric was thus placed in our newly introduced family of semi-metrics. The parameter $p$ sets up a trade-off of sorts: as $p$ gets larger, $d^p_{MJ}$ becomes closer to a metric satisfying the triangle inequality. However, as $p$ gets smaller, Proposition \ref{prop:Hausdorff} and Corollary \ref{cor:Hausdorff} show that $d^p_{MJ}$ is less affected by outlier elements.

It is with this prior literature and body of work in mind that, when we wish to measure discrepancy between time series' sets of structural breaks (which are finite sets), the MJ$_p$ semi-metric in (\ref{eq:MJdefn}) was a natural choice. As for the precise selection of $p$, our optimization framework did not particularly rely on the triangle inequality, while outlier sensitivity is much more important. Hence, we select a small value of $p$, in this case $p=\frac12.$

\subsection{Illustration study of different (semi-)metrics} 

\begin{figure}[H]
    \includegraphics[width=.97\textwidth]{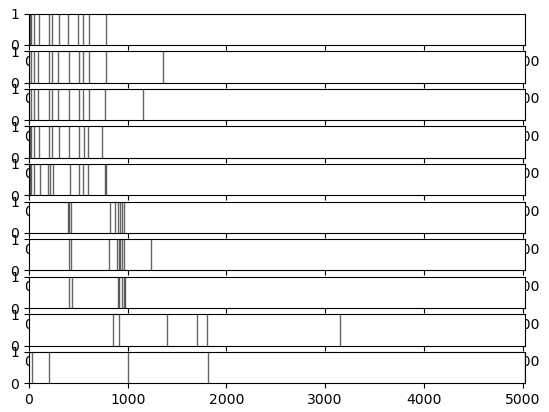}
    \caption{Collection of ten synthetic time series with structural breaks displayed. Two clusters of similarity (TS1--TS5 and TS6--TS8) are observed, as well as two outlier elements (TS9 and TS10) not similar to anything else.}
    \label{fig:TSoutliers}
\end{figure}

In this section, we generate some figures to graphically illustrate different values the aforementioned (semi-)metrics between sets can take. First, we generate a collection of ten time series each, inspired by the synthetic time series in Section \ref{sec:synthetic1} and experiments in \cite{James2020_nsm}. The collection is displayed in Figure \ref{fig:TSoutliers}, chosen to feature time series with moderate outlier elements. In this scenario, we consider the first five time series (TS1--TS5 inclusive) as similar, the next three (TS6--TS8) as similar, and  the final two (TS9 and TS10) as dissimilar to all other time series.

\begin{figure}[H]
    \begin{subfigure}[b]{0.49\textwidth}
        \includegraphics[width=\textwidth]{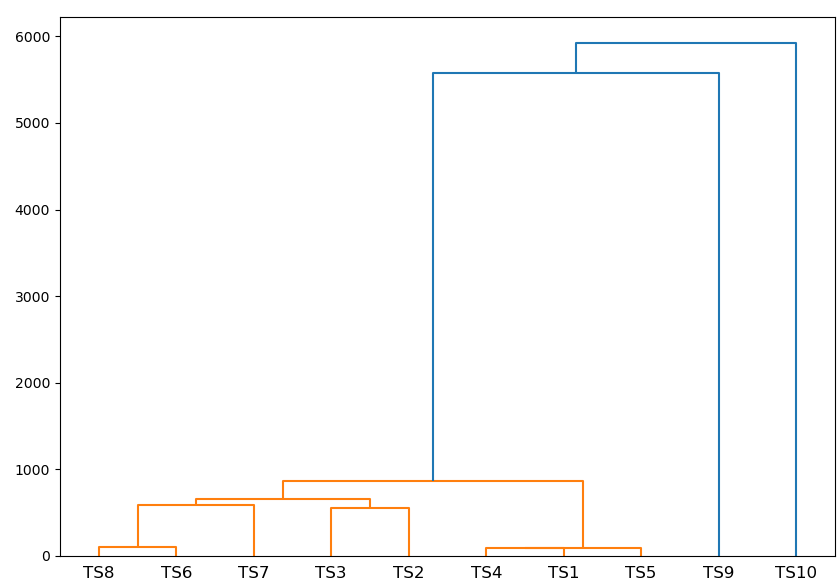}
        \caption{}
        \label{fig:moderate_haus}
    \end{subfigure}
    \begin{subfigure}[b]{0.49\textwidth}
        \includegraphics[width=\textwidth]{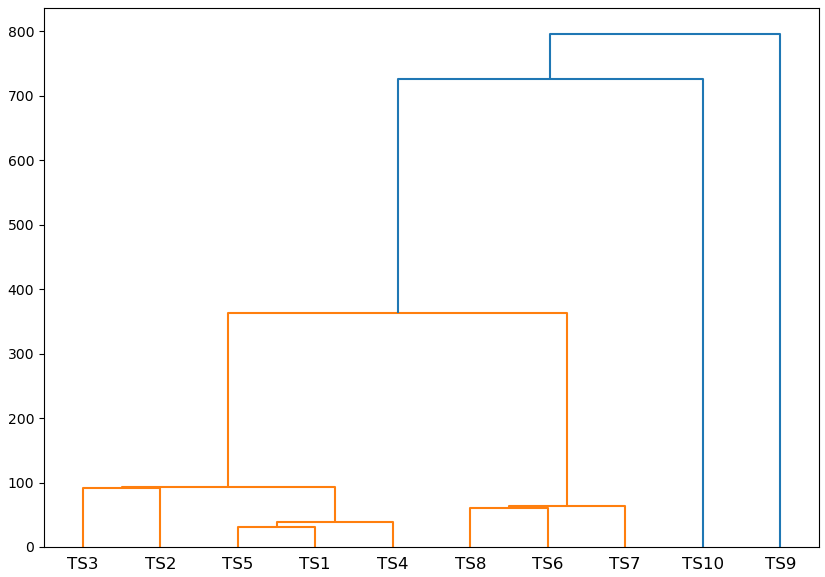}
        \caption{}
        \label{fig:moderate_mj05}
    \end{subfigure}\vspace{3pt}
    
        \begin{subfigure}[b]{0.49\textwidth}
        \includegraphics[width=\textwidth]{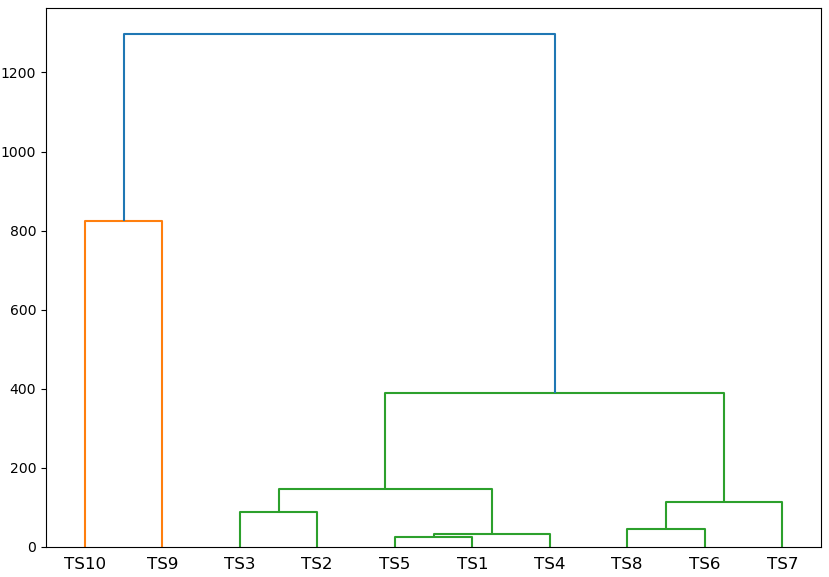}
        \caption{}
        \label{fig:moderate_mj1}
    \end{subfigure}
        \begin{subfigure}[b]{0.49\textwidth}
        \includegraphics[width=\textwidth]{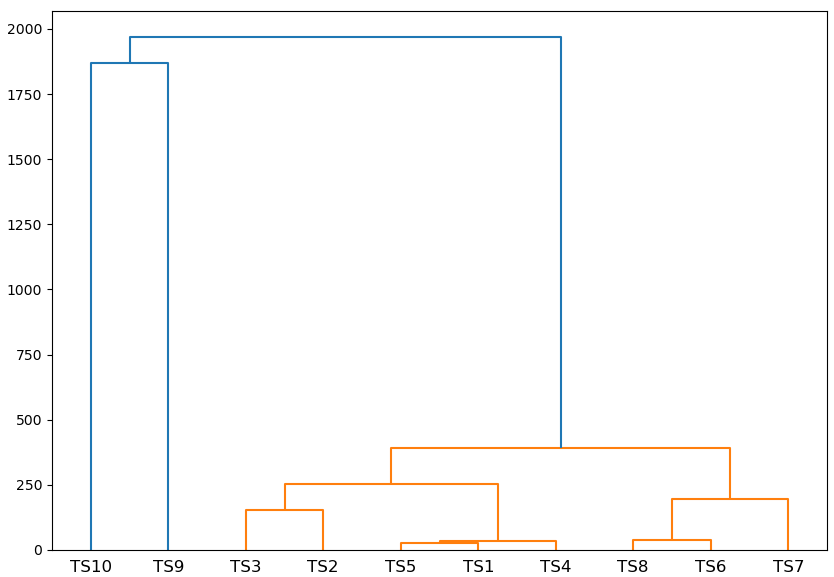}
        \caption{}
        \label{fig:moderate_mj2}
    \end{subfigure}
    \caption{Hierarchical clustering applied to the ten synthetic time series in Figure \ref{fig:TSoutliers} using (\textbf{a}) the Hausdorff metric, (\textbf{b}) the MJ$_{0.5}$, (\textbf{c}) the MJ$_1$, and (\textbf{d}) the MJ$_2$. Results indicate that the  MJ$_{0.5}$ does the best job at distinguishing between the two clusters of similarity and highlighting the dissimilarity of the two outliers.}
    \label{fig:dend_outliers_moderate}
\end{figure}

The graphical representation of these distances is supplied in Figure \ref{fig:dend_outliers_moderate}, in which we apply hierarchical clustering to the collection of synthetic time series, using the Hausdorff, MJ$_1$, MJ$_{0.5}$, and MJ$_2$ distances. Even in this instance of moderate outliers, the Hausdorff distance (\ref{fig:moderate_haus}) fails to correctly identify the general structure in the time series collection (one cluster of 1--5, one of 6--8, and two outliers). The remaining three semi-metrics correctly identify the general structure in the time series collection. As predicted, the MJ$_{0.5}$ (\ref{fig:moderate_mj05}) does the best job, both at distinguishing between the two clusters of similarity as well as highlighting the fact that the two outliers are distinct from everything else. We thus provide this as graphical evidence of the suitability of $p=\frac12$ to handle outlier elements, a necessary part of our overall optimization framework, as mentioned first in Remark \ref{remark:outliers}.

\begin{adjustwidth}{-\extralength}{0cm}

\reftitle{References}



\end{adjustwidth}
\end{document}